\documentclass[final,authoryear,5p,times]{elsarticle}

\usepackage{amssymb}
\usepackage{url}
\usepackage{longtable}
\usepackage{siunitx}
\usepackage{array}
\usepackage{multicol}
\usepackage{todonotes}
\usepackage{float}
\restylefloat{table}

\newcolumntype{L}[1]{>{\raggedright\let\newline\\\arraybackslash\hspace{0pt}}m{#1}}
\newcolumntype{C}[1]{>{\centering\let\newline\\\arraybackslash\hspace{0pt}}m{#1}}
\newcolumntype{R}[1]{>{\raggedleft\let\newline\\\arraybackslash\hspace{0pt}}m{#1}}

\journal{Astronomy and Computing}

\begin{document}

\begin{frontmatter}



\title{A BOINC\tnoteref{boinc} based, citizen-science project for pixel Spectral Energy Distribution fitting of resolved galaxies in multi-wavelength surveys}
\tnotetext[boinc]{Berkeley Open Infrastructure for Network Computing}


\author[kv]{Kevin~Vinsen\corref{cor1}}
\ead{kevin.vinsen@icrar.org}
\cortext[cor1]{Corresponding Author}
\address[kv]{International Centre for Radio Astronomy Research (ICRAR),\\
The University of Western Australia,\\
M468, 35 Stirling Highway, Crawley, Perth, WA 6009, Australia }

\author[dt]{David Thilker}
\address[dt]{Department of Physics \& Astronomy, \\
The Johns Hopkins University, \\
3400 N. Charles Street, \\
Baltimore, MD 21218, USA}

\begin{abstract}
In this work we present our experience from the first year of theSkyNet Pan-STARRS1 Optical Galaxy Survey (POGS) project.
This citizen-scientist driven research project uses the Berkeley Open Infrastructure for Network Computing (BOINC) middleware and thousands of Internet-connected computers to measure the resolved galactic structural properties of $\sim$100,000 low redshift galaxies. 
We are combining the spectral coverage of GALEX, Pan-STARRS1, SDSS, and WISE to generate a value-added, multi-wavelength UV-optical-NIR galaxy atlas for the nearby Universe. 
Specifically, we are measuring physical parameters (such as local stellar mass, star formation rate, and first-order star formation history) on a resolved pixel-by-pixel basis using spectral energy distribution (SED) fitting techniques in a distributed computing mode. 
\end{abstract}

\begin{keyword}

BOINC \sep Pan-STARRS1 \sep Cloud Computing \sep Spectral Energy Distribution \sep Distributed Client Computing \sep MAGPHYS
\end{keyword}

\end{frontmatter}


\section{Introduction}

TheSkyNet \citep{theSkyNet} is a citizen science project that aims to raise the public profile of science, specifically astronomy and radio astronomy.
At the same time it creates a research grade data processing resource for astronomers. 
TheSkyNet allows members of the public to donate their spare computing power to process astronomy data, with the software running quietly in the background of their personal machines.

In August 2012, The International Centre for Radio Astronomy Research (ICRAR) and The Johns Hopkins University (JHU) signed a Memorandum of Understanding for theSkyNet to process observations of galaxies observed by Pan-STARRS1 (PS1) \citep{doi:10.1117/12.859188}.
This dataset is complemented by multi-wavelength imaging from: the Sloan Digital Sky Survey (SDSS) \citep{York:2000vr}, Galaxy Evolution Explorer (GALEX) \citep{Martin:2005ko} and Wide-field Infrared Survey Explorer (WISE) \citep{2010AJ....140.1868W}.

The Panoramic Survey Telescope and Rapid Response System (Pan-STARRS) will use gigapixel mosaic cameras on four modest aperture telescopes to survey the entire Hawaiian sky in visible and near-IR bands. 
The first Pan-STARRS prototype survey began in May 2010 using a single telescope system (PS1).  
The PS1 telescope has a 1.8-m f/4 design, producing a 3$^\circ$ diameter field of view. 
Images are recorded on a 1.4 gigapixel CCD camera \citep{2009amos.confE..40T}, the worldÕs largest instrument of its kind. 
All PS1 science goals are supported by several parallel surveys \citep{2006amos.confE..39C}.  
The $3\pi$ Steradian Survey covers the entire sky north of declination -30$^\circ$ with a total of 60 epochs, twelve in each of 5 filters (\textit{grizy}).   
The $3\pi$ survey is providing images of higher resolution (mode FWHM = 1.2" for \textit{g} and \textless1" for all other bands) and moderately higher sensitivity than SDSS, over a much larger area on the sky.

TheSkyNet Pan-STARRS1 Optical Galaxy Survey (POGS) project \citep{pogs1} was created to efficiently process and interpret the multi-wavelength data from the resolved galaxies detected by PS1. 
This large volume of data would take many tens of years to process on a large University computing cluster.
We believe this can be done in 2-3 years with theSkyNet using the BOINC middleware \citep{boinc1}. 
For each catalogued galaxy meeting our selection criteria, the computers of our volunteers are being harnessed, via the Internet, to perform pixel-by-pixel spectral energy distribution (SED) fitting.
This will provide us with physical parameters about the galaxy, such as: the stellar mass, star formation rate (SFR), dust attenuation, and first-order star formation history (SFH).
With the resultant pixel SED fitting output, we will then constrain parametric models of galaxy morphology in a more meaningful way than ordinarily achieved. 
In particular, we will fit multi-component (bulge, bar, disk) trial galaxy models, directly to the distribution of stellar mass rather than surface brightness in a single band, which is locally biased by star formation activity and dust obscuration. 

\newcommand{\asOfDate}{30-Sep-2013}
To date (as at \asOfDate) the project has 4,663 active users with 11,629 computers; and has processed 8,348 galaxies or 53 million pixels.  
We are presently a modestly sized BOINC project compared to the `@Home' projects, which have been operating for many years and have tens of thousands of volunteers, and in the case of SETI@Home, a million plus volunteers.
We anticipate significant continued growth in our user base as we are currently growing at $\sim$50-60 computers a day.

The acronym POGS is a reference to a game played with discs, originating on Maui, Hawaii, in the 1920s \citep{pogs2}; and the fact that the Pan-STARRS PS1 telescope, is situated on Mount Haleakala, Maui.

\section{Spectral energy distribution fitting \& MAGPHYS}
\label{Spectral energy distribution fitting}
The power of multi-wavelength data for understanding the substructure of nearby galaxies has only just begun to be exploited systematically. 
Various authors have described methods to extract the information content in a pixel using SED based methods. 
These include:
\begin{enumerate}
\item SED fitting \citep{Crockett:2011is, Jeong:2007ik};
\item pixel-z \citep{Conti:2003jr, Welikala:2009kp, Wijesinghe:2010fr};
\item PCA \citep{Zibetti:2011br}. 
\end{enumerate}

For this project, the pixel-by-pixel SED fitting is done using the `Multi-wavelength Analysis of Galaxy Physical Properties' code (MAGPHYS) of \citet{daCunha:2008cy}. 
We have adapted it to run within the BOINC distributed-computing framework. 
Other SED fitting codes are also being evaluated for our project.
MAGPHYS was implemented first because it offers a physically motivated model to interpret stellar and dust emission at UV, optical, and and infrared wavelengths in a self-consistent manner. 

Evolution of the intrinsic starlight from stellar populations of a specified SFH are computed using the \citet{Bruzual:2003ck} population synthesis code.  
The allowed SFHs are realistic in the sense that they are a combination of parameterised exponential declines plus a superposition of random bursts at later times.

MAGPHYS adopts the two-component model of \citet{Charlot:2000bd} in order to realistically predict the attenuation of starlight for stars of various ages, and the corresponding re-emission of this energy by dust in the interstellar medium (ISM). 
The two redshift-dependent model libraries consist of: one library of 50,000 combinations of stellar population spectra, reflecting the SFH (UV-optical) model being used; and a second library to describe the Infrared emission from dust, using the approach of \citet{daCunha:2008cy}. 

\begin{figure} [hbt]
\centering
\fbox{\includegraphics [clip, width=0.45\textwidth] {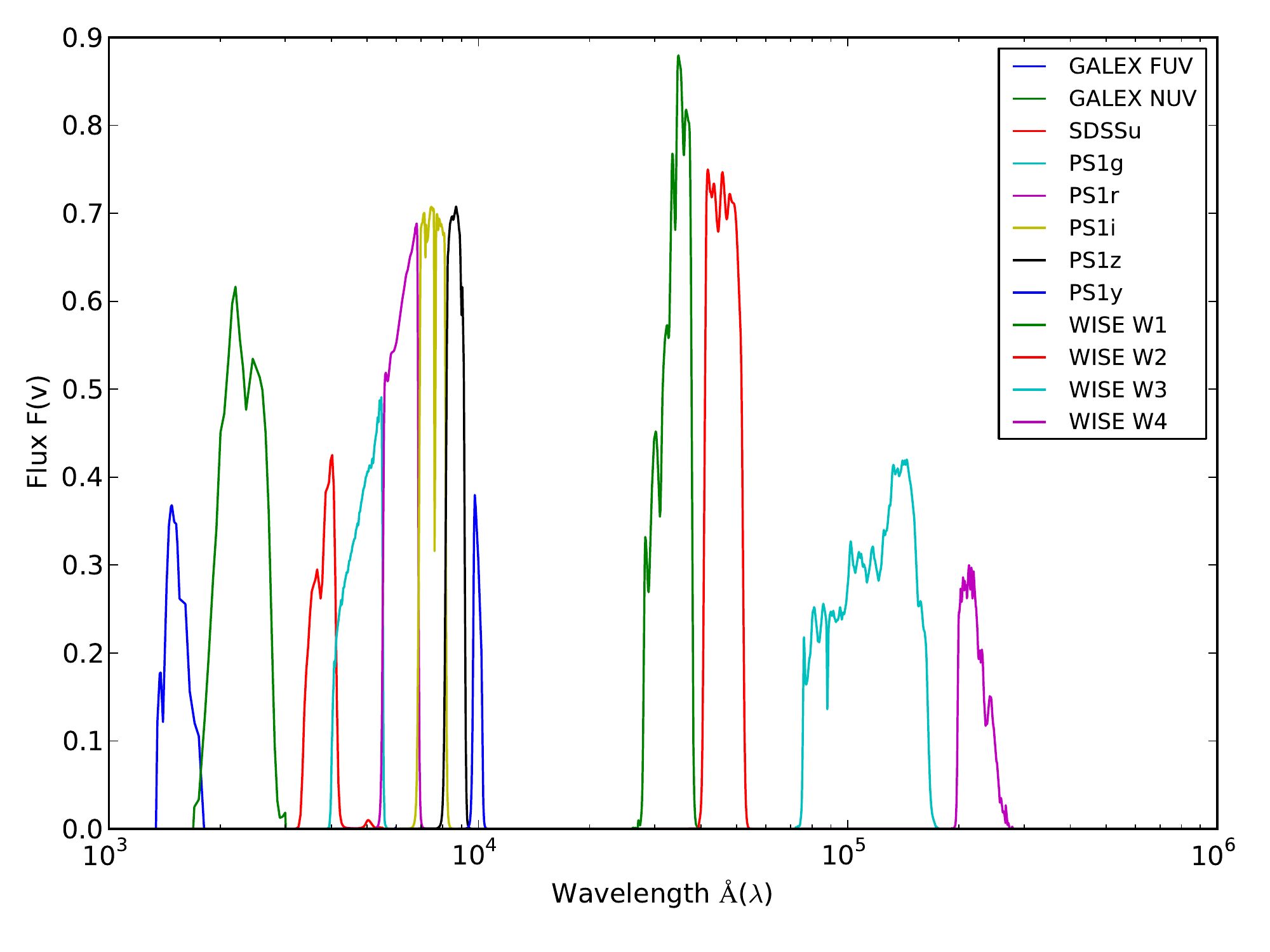}}
\caption{Typical filter curves used in a MAGPHYS run with PS1 data.}
\label{fig:filters}
\end{figure}

From these 1.54GB files, MAGPHYS generates two models to be used for the SED fitting based on the filters used in the observations.
The filter bands used in theSkyNet POGS at the time of writing are shown in Figure \ref{fig:filters}.
MAGPHYS allows the redshift of these model files to be within $z=0.005$ of the redshift of the observed pixels. 
This means we do not have to generate model files for every redshift of every galaxy. 
Currently, the observed galaxies have varying redshifts up to a maximum $z=0.1$, so we have created model files for redshifts of: 0.00, 0.01, 0.02, \ldots, 0.09, 0.10.  

The resultant model files are quite large, and increase in size with the number of filter bands being processed. 
Using the filter bands listed in Figure \ref{fig:filters} the star formation model files (describing the UV-optical portion of the SED) are $\sim$15-20MB, and the infrared model files (describing the dust properties) are $\sim$5-10MB.

Using these two models, MAGPHYS takes the observed fluxes and the associated uncertainties through the various filters (both in Jy), and calculates the physical parameters shown in Table \ref{tab:magphys}.  
For each parameter the seven values listed in Table \ref{tab:magphys-layers} are recorded.
The likelihood distributions for each of the physical parameters in Table \ref{tab:magphys} are also recorded.

\section{BOINC distribution \& processing}
\label{BOINC distribution & processing}
The BOINC middleware is well proven through projects like: SETI@Home \citep{Korpela:2012jt} and Einstein@Home \citep{Allen:2013eb}.
A detailed description of the BOINC middleware can be found in \citet{Korpela:2012jt} and \citet{Allen:2013eb}.
We will only provide a brief overview here.
The basic components of a BOINC server are:
\begin{enumerate}
\item the BOINC daemons, called the scheduler, feeder, transitioner, validator, assimilator and file deleter.
\item the Apache web server with two CGI programs.
\item the MySQL database server.
\end{enumerate}

Only two programs must the provided by the project: the validator and the assimilator.
The rest are provided `out of the box' by the BOINC project.

The validator decides whether results returned by the clients are correct.
It does this by comparing the results of runs from different computers.
If the results match it assumes the results are valid.
The MAGPHYS code rounds its output to only 3 decimal places, which means results tend to be consistent across different operating systems and hardware. 
Differences do occasionally occur, mainly when an old 32-bit machine compares results against a 64-bit machine or an Android device.
This is discussed further in Section \ref{Performance}.
TheSkyNet POGS validator performs a simple line by line comparison.
It excludes the end of line characters because they are different for different operating systems.
If the validator cannot get a matching result after five attempts it will abandon that particular work unit.

The assimilator processes completed jobs. 
It parses the text file output from MAGPHYS and only stores the 4 results needed to draw the images described in Section \ref{Web Site} into the database.
The text file is then copied to a high availability file store for processing when all the pixels for a galaxy have been returned.

\begin{figure*} [!t]
\centering
\fbox{\includegraphics [clip, width=0.80\textwidth] {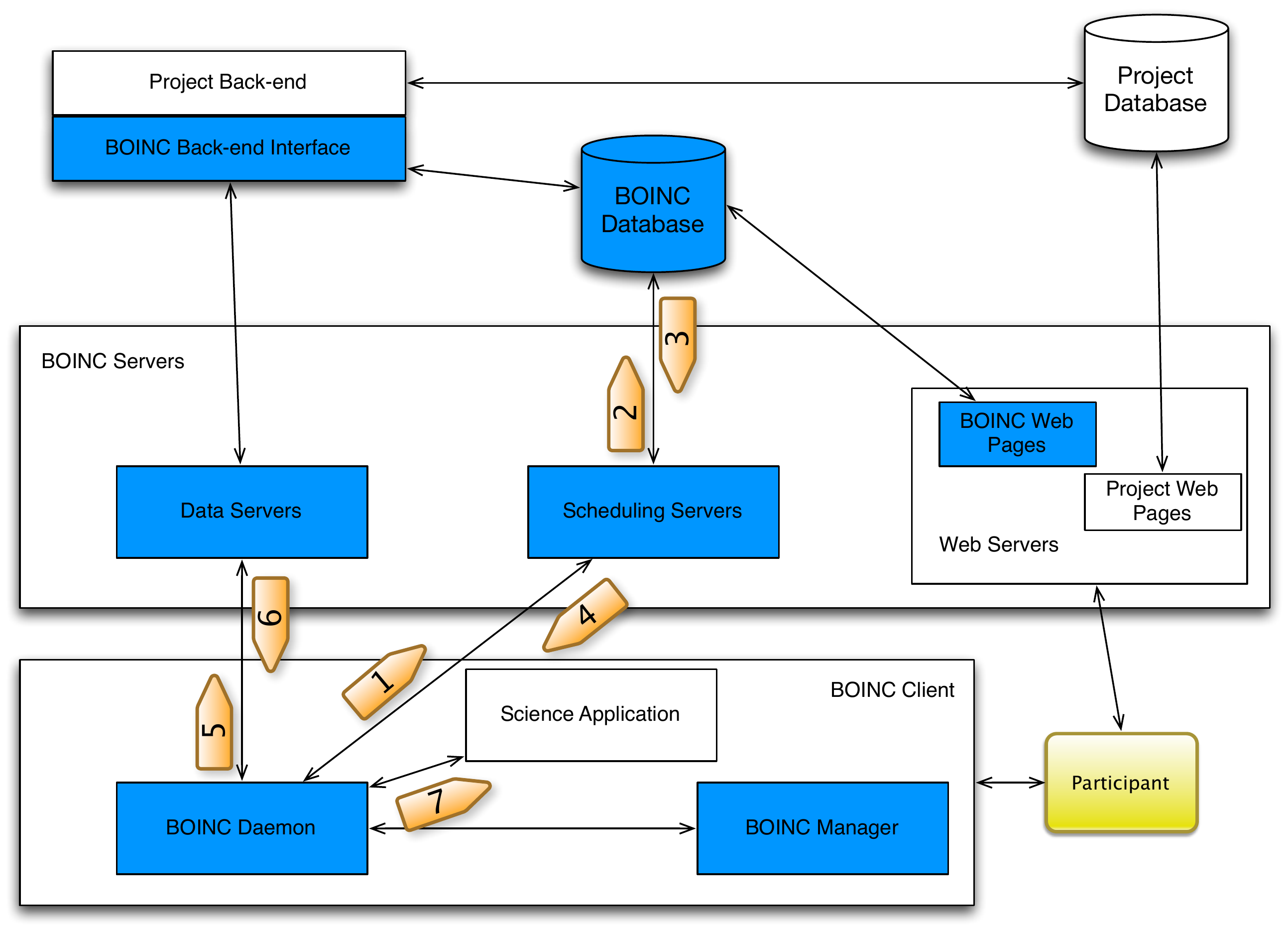}}
\caption{The data flow for a BOINC client to request data from a BOINC server.}
\label{fig:BOINCArchitecture01}
\vspace{5mm} 
\fbox{\includegraphics [clip, width=0.80\textwidth] {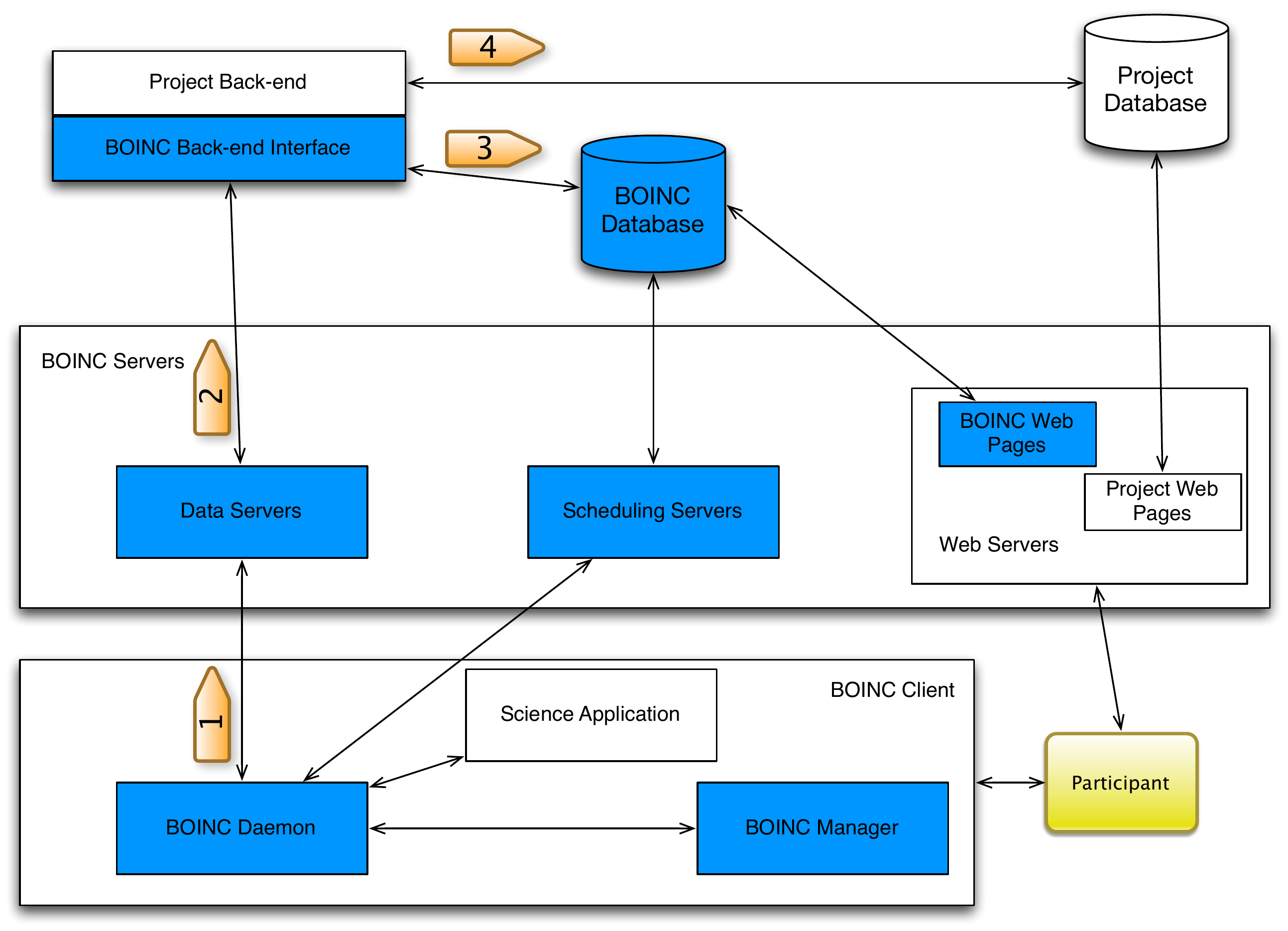}}
\caption{The data flow for a BOINC client to upload results to a BOINC server.}
\label{fig:BOINCArchitecture02}
\end{figure*}

An overview of the process of getting work to the clients is given in Figure \ref{fig:BOINCArchitecture01}.
The steps are:
\begin{enumerate}
\item The BOINC client requests a number of new work units (a work unit describes the computation to be performed) from the BOINC server.
\item The server checks the database for work units. 
BOINC has a number of different distribution methods depending on the nature of the project. 
POGS uses the simplest priority-order, which takes high priority tasks first (created when validation fails) then tasks in the order they were created.
\item The database gives back details of work units to be assigned to the client.
\item The scheduling server passes the details of work units to be assigned to the client.
\item The client requests the work units from the data server.
\item The data server passes the work units to the client.
\item The client calls the science application (MAGPHYS) with the work unit details.
\end{enumerate}

The BOINC middleware allows the client to create a queue of work units for processing later. 
This means the client does not need to be online all the time and can process data offline.

Once the client has data to return to the server the process shown in Figure \ref{fig:BOINCArchitecture02} is followed.
The client can return the results of many work units in one transfer.
The steps are:
\begin{enumerate}
\item The results are stored on the data server.
\item The data server informs the back-end that data is available.
\item The validator checks the results by comparing them with the results from another client. 
If the results match it is made available to the assimilator, and the users receive their credit for processing the work unit.
If the results do not match the work unit is pushed out to another client to see if another client can produce the same results.
\item The assimilator parses and stores some of the output from MAGPHYS and copies the file to a high availability file store.
\end{enumerate}

\section{Data Pipeline}
\label{Data Pipeline}
\subsection{Image Preparation}

Image preparation is performed at JHU.
The exact details of the image preparation are beyond the scope of this paper and will be published separately (Thilker et al., in prep.).
Galaxy images are extracted from the PS1 datastore at the Institute for Astronomy (IfA, Monoa, HI). 
GALEX, SDSS and WISE data is then added to produce two multi-extension FITS \citep{fits1} files.
All multi-wavelength images are convolved to a common angular resolution and placed on a consistent flux scale (Jy/arcsec$^2$).
The first FITS file contains the observed fluxes, and the second contains the associated uncertainties for the observed fluxes. 
The FITS files have a layer for each of the filter bands there is data for. 
As part of the FITS header each layer contains information giving the name and the effective wavelength of the filter.

\subsection{Model File Preparation}

Not all the Pan-STARRS galaxies have associated infra-red and ultraviolet data that can be used.
This can be due to factors such as:
\begin{enumerate}
\item the archival coverage in the UV;
\item the UV and IR data does not have sufficient resolution to add significant value on substructure for objects which have a small angular size.
\end{enumerate}

For a given batch of galaxies, with the same filter bands, we generate the model files as described in Section \ref{Spectral energy distribution fitting} with a limited set of filter bands for that run.
This reduces the processing time for each pixel and the amount of data that has to be downloaded to the client.

The resulting model files are loaded onto the main BOINC server.
This is a one-off process as tens of thousands of galaxies can be processed from one set of model files.
BOINC allows us to mark the model files as `sticky', so that once it is downloaded it stays on the client indefinitely or until BOINC issues a command to flush it from the client's disk.

\subsection{Queueing of galaxies}
\label{Queueing of galaxies}
The FITS files containing the image data and associated errors for a galaxy are loaded onto the main BOINC server and the galaxy is registered in the database.
Every 10 minutes, a BOINC task is triggered which checks the BOINC queues to see how many BOINC work units are available.
If a predefined threshold has been passed, the system takes the next available galaxy to produce new BOINC work units.
The number of work units generated by a galaxy is dependent on the size of the galaxy and the sparseness of data in the FITS files and across layers.
The FITS files are opened and many MAGPHYS observation files are created. 
Each observation file is stored and is linked to a BOINC work unit.
The work unit is then queued to be sent to two clients initially.

To give the BOINC community more sense of involvement in the POGS project, we created an online feature to show the users what areas of a galaxy they have worked on.
This allows the users to know exactly where their CPU cycles are going. 
This meant we needed to give the user areas of the image to process rather than individual pixels.

\begin{figure}  [hbtp]
\centering
\fbox{\includegraphics [clip, width=0.45\textwidth] {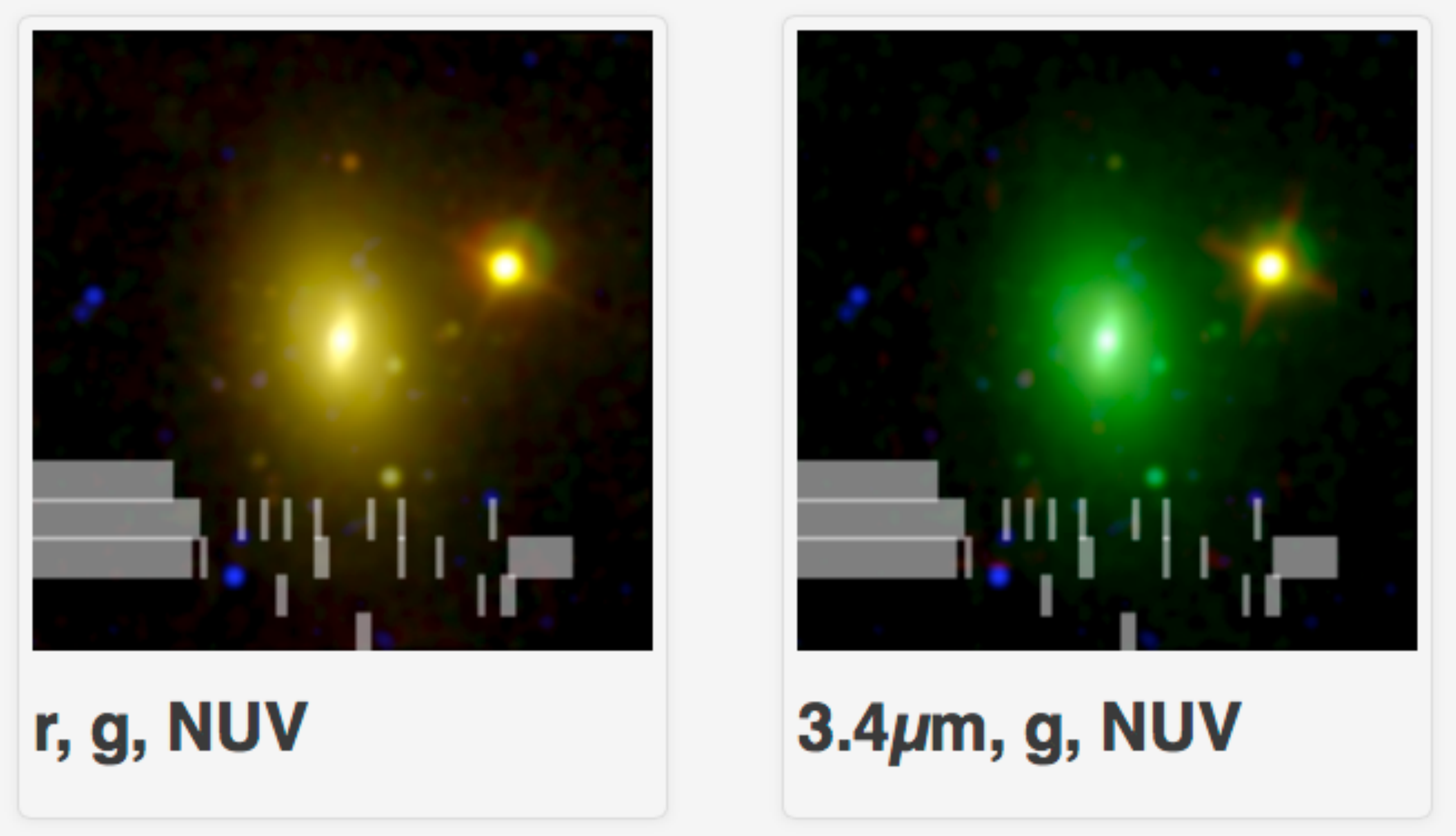}}
\caption{The POGS screen showing where the user has worked NGC4421.}
\label{fig:pogsImage1}
\end{figure}

A line of pixels in a typical galaxy image having width of several hundred to several thousand pixels is very hard to see, but an area 7 pixels high by 3 or 4 wide is more readily visible as shown in Figure \ref{fig:pogsImage1}.
To generate this the work unit generator starts with a 7 pixel high slice of the image.
It counts the number of useable pixels, and if the grand total is less than a configurable number it gets the next slice.
A useable pixel is one that has data for at least 6 frequency bands.
All the operations are configurable using a simple configuration file. 

In Figure \ref{fig:pogsImage1} the grey areas are where the user has processed pixels for this galaxy.
The 3 larger rectangle on the left of the figure only had a few useable pixels, it took many slices to find enough useable pixels to compile an area.
Whereas the smaller rectangles near the centre of the image were full of useable pixels.

\subsection{Archiving}
\label{Archiving}

Once all the pixels in a galaxy have been processed the results need to be collated ready for storage.  
The collated data needs to be stored in a long term archive, such as: Amazon Web Services (AWS) Simple Storage Service (S3), AWS Glacier, and the Pawsey Centre Hierarchical Storage Manager run by iVEC \citep{ivec1}.
The data does not need to be available on disk 24x7.
Some lag whilst a file is retrieved from tape is acceptable.

Every four hours a BOINC task is activated to check if all the work units for a galaxy have been returned.
When there are no more results outstanding the galaxy is marked for archiving.
The system then collates the data into an HDF5 file and moves the resulting file to long term storage.
It is first copied to an S3 bucket, then it is moved to the Next Generation Archiving System (NGAS) \citep{2012SPIE.8451E..18W} at the Pawsey Centre in Perth, Western Australia (WA).

Within S3 it is possible to define a `lifecycle'.
This is a mechanism to move the data to Glacier after a period of time to reduce costs.
The penalty is a 3-4 hour access time. 

\subsection{BOINC Tasks}
\label{BOINC Tasks}
The BOINC middleware comes with a simple task scheduling system.
We use this for generating new work units and collating the data once all the pixels of a galaxy are complete.
Table \ref{tab:boinc-tasks} shows the major tasks that we developed to enable the smooth running of the system with minimal operator intervention.

\begin{table} [hbt]
\centering
\begin{tabular}{ | L{0.09\textwidth} | C{0.06\textwidth} | L{0.25\textwidth} | }
\hline
Task & Period & Description of BOINC task \\
\hline
\hline
FITS to work unit & 10 mins & Check the queue of work units. If it has fallen below a threshold generate more work units. \\
\hline
assign credit & 24 hours & Ensure the users are credited for the areas of a galaxy they worked on. \\
\hline
build PNG images & 3 hours & Build the PNG images that show the latest $\mu$, $M_*$, $L_d^{tot}$ and $\psi(t)$ maps. \\
\hline
delete galaxy & 4 hours & Once a galaxy has been archived after N days delete the database entries. \\
\hline
archive HDF5 galaxy & 4 hours & Once all the pixels for are galaxy are completed generate the HDF5 file. \\
\hline
processed galaxies & 4 hours & Check to see if all the pixels for a galaxy are completed. \\
\hline
store files & 4 hours & Move the HDF5 file to long term high availability storage. \\
\hline
\end{tabular}
\caption{The BOINC tasks developed for theSkyNet POGS.}
\label{tab:boinc-tasks}
\end{table}

The BOINC project also provides tasks \citep{boinc3} to perform house-keeping operations.
These tasks are to keep a BOINC based project operating efficiently such as: output statistics, importing BOINC teams, updating the user of the day, database archiving.   

\section{Infrastructure}
\label{Infrastructure}
\subsection{Amazon Web Services} 
\label{Amazon Web Services} 
The server side computing resources are provided through Amazon Web Services.
This was initially supported by an `AWS in Education grant award' of US\$2,000 and was the main motivation for choosing AWS initially.

\begin{figure*} [hbt]
\centering 
\fbox{\includegraphics [trim=13mm 3mm 12mm 16mm, clip, width=0.85\textwidth] {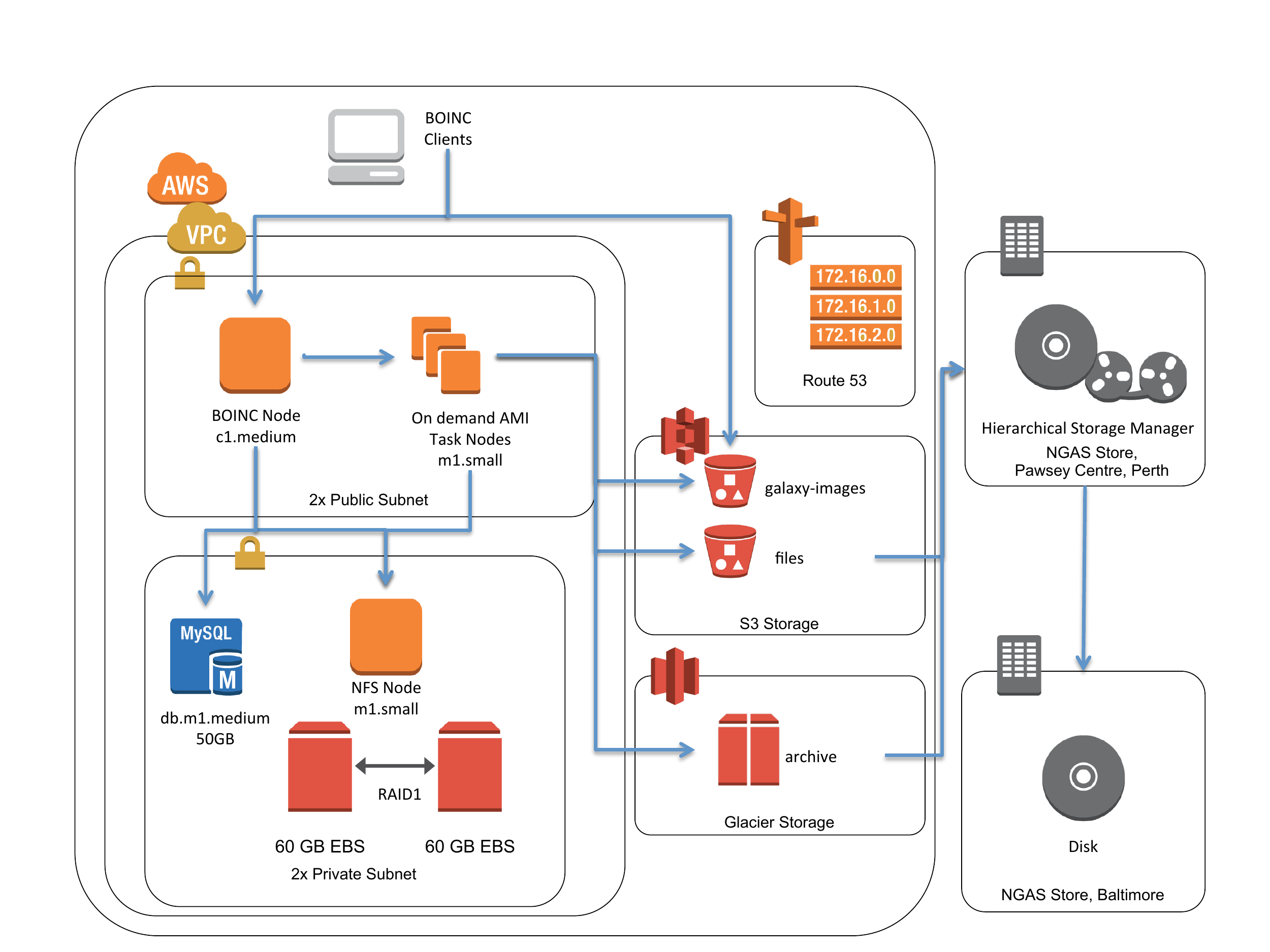}}
\caption{The V2 POGS Architecture.
A Virtual Private Cloud (2 public and 2 private subnets each in a different availability zone), 1x c1.medium instance, 1x m1.small instance, 1x db.m1.medium instance, on-demand m1.small instances, Route 53, S3 and Glacier storage, and 2 NGAS archive servers}
\label{fig:pogs2}
\end{figure*}

The first version of the infrastructure (see Section \ref{Comparison of V2 and V1 Infrastructure}) did not use many of the AWS specific technologies.
At the start of the project we were undecided if we would stay with AWS, and wanted to allow a relatively easy migration path once the grant had been used up to either bespoke hardware or a different cloud provider.
When we had a stable functioning system and had spent $\sim$60\% of the grant (December 2012), we performed a simple cost benefit analysis.    
Two major factors contributed to the project staying with AWS and implementing version 2 of the infrastructure.
These were:
\begin{enumerate}
\item Ease of use. 
AWS provides a very simple web interface for developers and there are libraries for Python, Ruby and C.
The AWS developer has complete control over the instances deployed in the cloud, without having to go through `institutional' administrators.
\item Cost and time of acquiring infrastructure.
The cost of running within the AWS Cloud was cheaper than buying the hardware and running it.
If the expected growth in 2013 started to happen we would not need to purchase more equipment, we could simply increase the power of our instances or start new instances as we need them.
\end{enumerate}

Version 2 has made use of more of the AWS features and significant improved the overall performance of the system and reduced the monthly costs.
This is discussed in more detail in Section \ref{Comparison of V2 and V1 Infrastructure}.

\subsection{BOINC Servers}
BOINC requires one node to be designated as the primary node.
This node is responsible for co-ordinating what daemons and tasks run on what machines, and when they run.
This is done through a single xml file called \textit{config.xml}.

To horizontally scale a BOINC based project it is necessary to have a shared file system which all the nodes can access \citep{boinc2}, and the key users having the same UID and GID.
BOINC then uses ssh to remotely execute daemons/tasks on the appropriate servers as required. 
An overview of the V2 infrastructure is shown in Figure \ref{fig:pogs2}.

To provide a shared file system we use a single Elastic Compute Cloud (EC2) instance with two 60GB Elastic Block Storage (EBS) volumes attached and configured as a RAID1 volume.
RAID1 is used to provide redundancy, because EBS volumes do fail, as will be discussed in Section \ref{Comparison of V2 and V1 Infrastructure}.
The instance runs an NFS server that only systems in the Virtual Private Cloud (VPC) can access. 
Snapshots of all the disks are taken twice daily to ensure that at most only 12 hours of data would be lost in the event of a catastrophic crash of AWS.
Should the NFS EC2 instance fail we can detach the EBS volumes and start a new one.

\begin{figure}  [hbtp]
\centering 
\fbox{\includegraphics [trim=4mm 5mm 5mm 0mm, clip, width=0.45\textwidth] {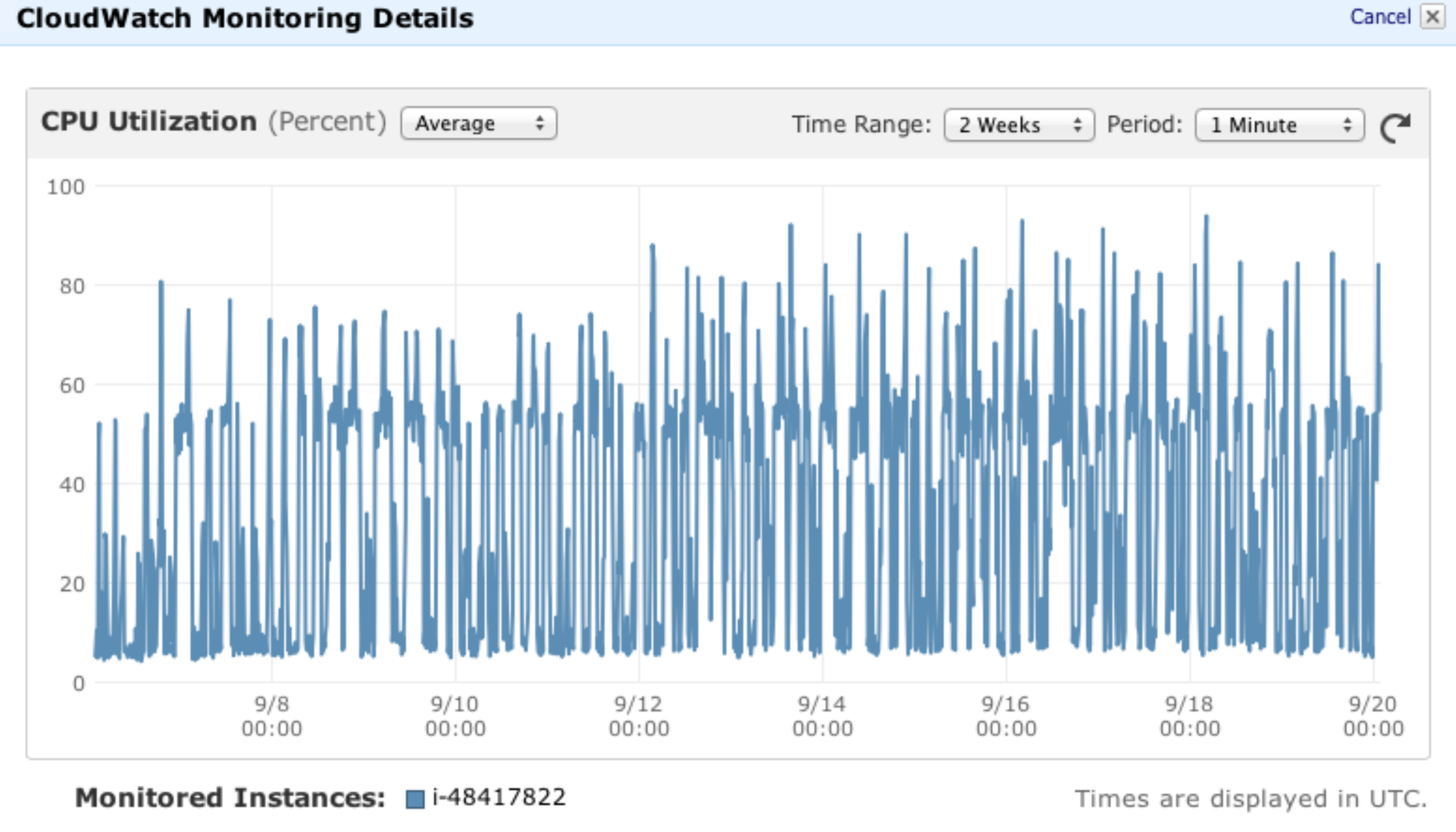}}
\caption{The current CPU load of the main server}
\label{fig:currentcpu}
\end{figure}

After the re-architecting from V1 to V2, all the daemons can all be run on a single modest instance.
The existing load is shown in Figure \ref{fig:currentcpu}.
Some short running tasks are still run on the main server (as can be seen by the short spikes), whilst longer running tasks are run on EC2 instances that are created just for that task from an Amazon Machine Image (AMI).

It must be remembered that Amazon bills per hour or part of thereof. 
So 5 minutes usage of a EC2 instance is still billed for 1 hour.
For the longer running tasks which may only be required 6 times a day this means we only pay for 6-8 hours of EC2 instance time instead of 24 hours.
For the smaller tasks it is not really economical to run them in their own instance.

If we find we need more CPU or Memory, it is a simple task to increase the capacity of the main server or to create a second or third EC2 instance to handle the load.

\subsection{Database}

The AWS Relational Database Service (RDS) used is based on the MySQL (5.6.12) database engine.
It is a db.m1.medium instance with $\sim$50GB of space assigned to it.
The database only records the minimum required to:
\begin{enumerate}
\item run the BOINC software.
\item to display false colour images to the BOINC users showing which areas they have been processing
\end{enumerate}

A snapshot of the database is taken once a day to allow for data recovery.
RDS does allow the user to create a read replica of the database in a different availability zone.
As load increases this option will be activated.

\subsection{Long Term Storage}
\subsubsection{HDF5}
\label{hdf5}

The text files produced by MAGPHYS (Section \ref{Spectral energy distribution fitting}) are large and irregular.
The likelihood distributions of the parameters listed in tables \ref{tab:magphys} and \ref{tab:magphys-layers} are variable length and quite detailed.
File formats such as FITS do not handle this type of data well.

\begin{figure}  [hbt]
\centering
\fbox{\includegraphics [clip, width=0.45\textwidth] {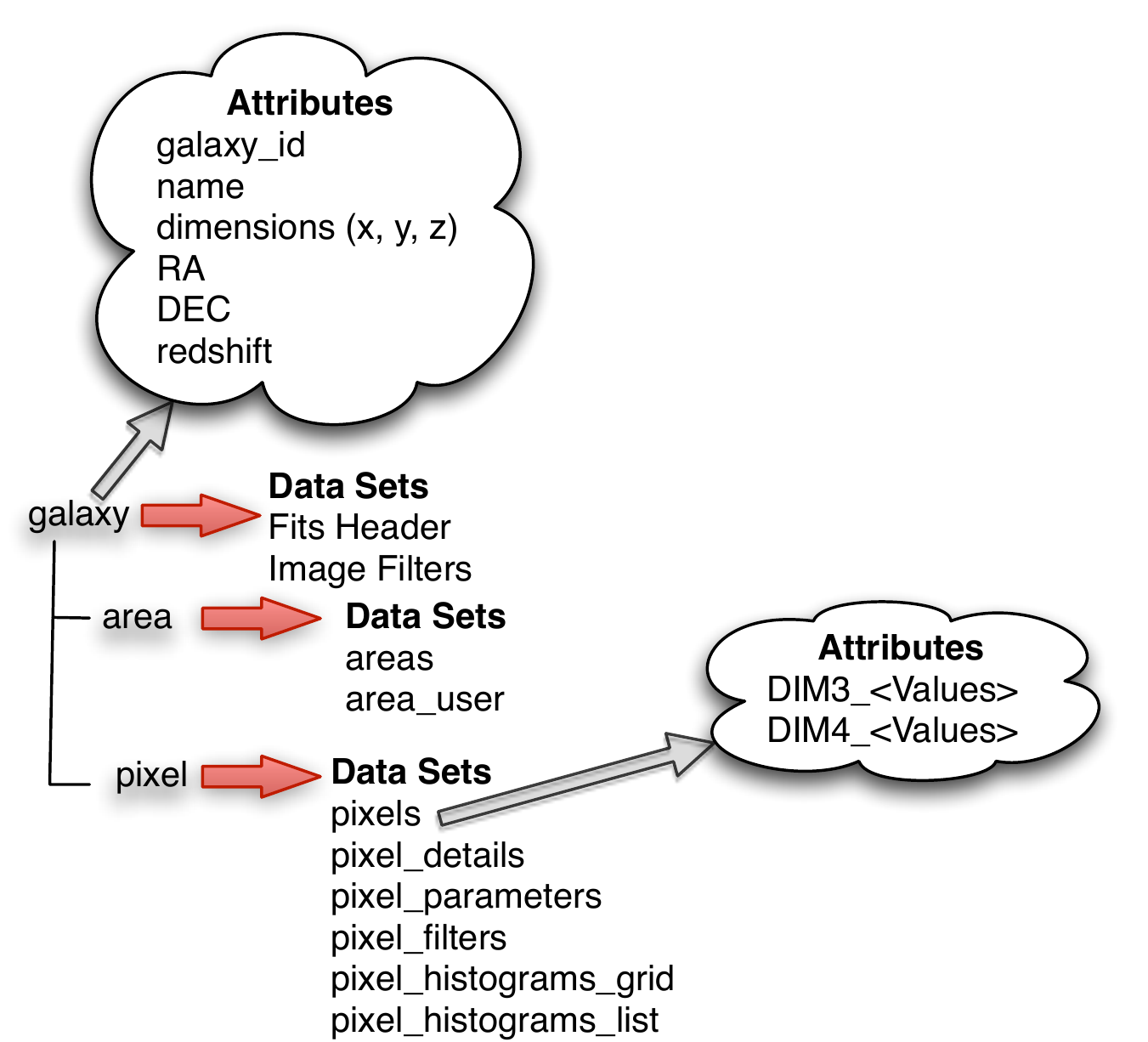}}
\caption{An overview of the HDF5 file format used.}
\label{fig:hdf5}
\end{figure}

Hierarchical Data Format (HDF, HDF4, or HDF5) is the name of a set of file formats and libraries designed to store and organise large amounts of numerical data.
The data is stored in groups and subgroups reminiscent of the POSIX filesystem.
Under a group any number of datasets can be stored and any number of attributes assigned to groups or datasets.
Figure \ref{fig:hdf5} shows an overview of the HDF5 structure used by POGS.

The main data set is the `galaxy/pixel/pixels', which is a number of 4 dimensional arrays of floating point numbers.
The dimensions are:
\begin{enumerate}
\item the X dimension from the original FITS file (DIM1)
\item the Y dimension from the original FITS file (DIM2)
\item the 16 parameters calculated by MAGPHYS as shown in Table \ref{tab:magphys} (DIM3)
\item the 7 values calculated for each parameter as shown in Table \ref{tab:magphys-layers} (DIM4)
\end{enumerate}

We have to tile the `galaxy/pixel/pixels' data because we face an increasingly common problem.
The output size for big galaxy images is too big to be held in memory, but the underlying software (numpy and h5py in this case) tries to create the whole image in memory.    

The likelihood distributions are stored in tiled one dimensional data sets in `pixel\_histograms\_list'.
These are then referenced from the X by Y grid `pixel\_histograms\_grid', which points to the start and end of the elements in the one dimensional data set.
All these elements are tiled to match the tiling of data in the `galaxy/pixel/pixels' group.

\begin{figure}  [hbt]
\centering
\fbox{\includegraphics [clip, width=0.45\textwidth] {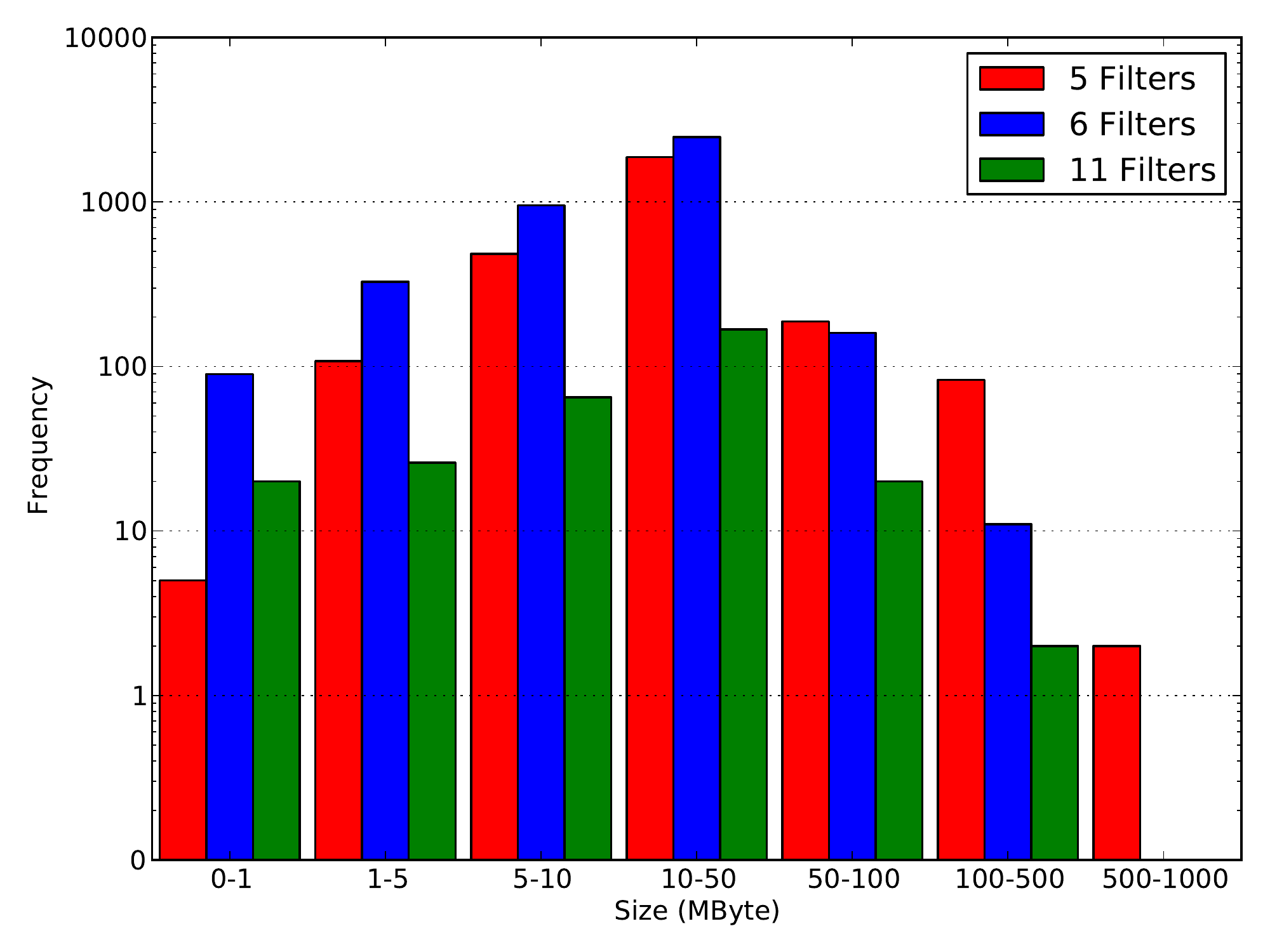}}
\caption{The HDF5 file sizes for various different filter combinations.}
\label{fig:file_sizes}
\end{figure}

Every four hours a BOINC task checks for galaxies that have been marked as complete.
It then generates a single HDF5 file from all the results that the assimilator has stored in S3.
This is then written to a bucket in S3.
The files range in size from a few MB up to just under 700MB as shown in Figure \ref{fig:file_sizes}.

Once the file is complete and safely stored in S3 it is transferred to the NGAS server at iVEC.
NGAS then archives the file and replicates it to other NGAS servers as necessary.

Once the HDF5 data has been stored in NGAS the space used in the database can be reclaimed.

\subsubsection{Next Generation Archiving System}
\label{NGAS}
NGAS was developed by the European Southern Observatory (ESO) to handle the large amount of astronomical data generated by its various telescopes.
The POGS project uses a modified versions of NGAS developed for The Murchison Wide Field Array (MWA) \citep{2012SPIE.8451E..18W}.

The NGAS MWA archive replicates data from Perth to Cambridge, Massachusetts. 
During Q4 of 2013 we hope to implement similar replication to move the final POGS data product from iVec in Perth, WA to JHU in Baltimore, Maryland.


\subsection{Comparison of V2 and V1 Infrastructure}
\label{Comparison of V2 and V1 Infrastructure}

\begin{figure}  [hbtp]
\centering 
\fbox{\includegraphics [trim=30mm 5mm 24mm 20mm, clip, width=0.45\textwidth] {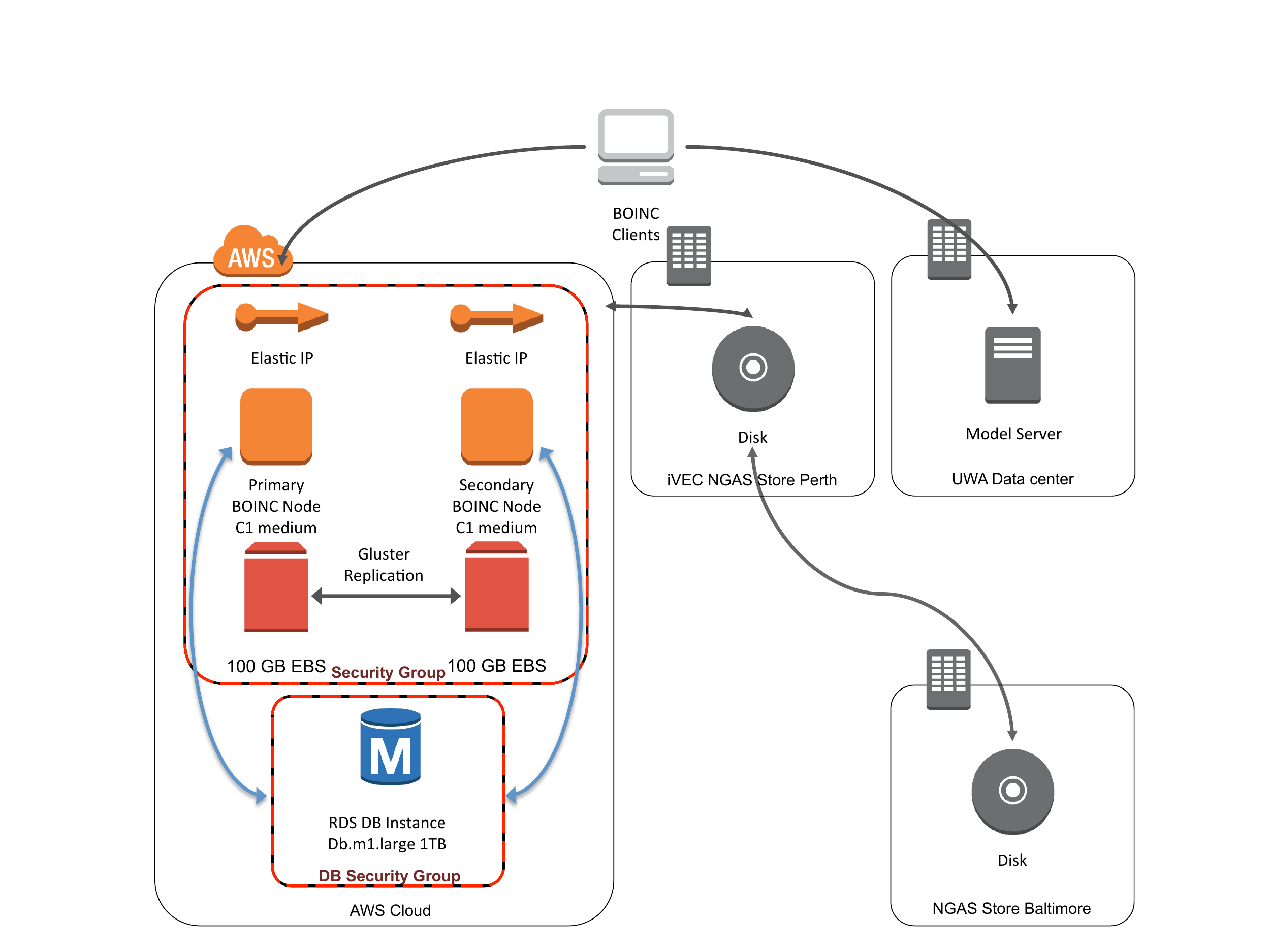}}
\caption{The V1 POGS Architecture.
2x AWS c1.medium instances, 1x AWS db.m1.large instance, 1 Model Server and 2 NGAS archive servers}
\label{fig:pogs1}
\end{figure}

As was discussed in Section \ref{Amazon Web Services} the V1 infrastructure was designed to be moved to different servers and infrastructure as easily as possible.
An overview of the V1 infrastructure is shown in Figure \ref{fig:pogs1}.

For V1 it was decided to assimilate and store the pixel data as quickly as possible by store them into an RDS MySQL database.
This resulted in us needing an AWS MySQL db.m1.large instance with $\sim$1TB of space assigned to it to store the data before it was archived to long term storage.
Because of the large amount of data to be stored this could take 20-30 seconds for a 21 pixel work unit.
Close to 96\% of the total database storage was used to store the likelihood distributions produced by MAGPHYS for each of the parameters described in Table \ref{tab:magphys}.
To manage the throughput we needed to run 6 assimilators across the two servers (4 on the primary node and 2 on the secondary node).

The database soon became the largest cost item in AWS.
The RDS price model includes paying for the storage in addition to the number of hours the database is running.
In the V2 version we only extract and store into the database the 4 values we need for the false colour ``progress'' images that are displayed on the web site.
The file returned by MAGPHYS is then stored on S3 ready to be processed at a later date.
Now a single assimilator is running and is able to assimilate a 21 pixel work unit in $\sim$0.35 seconds.
Our bill has dropped from $\sim$US\$800 to $\sim$US\$450 a month.
We will be buying reserved instances in the next few months once the V2 system is completely `bedded in' which will reduce the monthly bill by another 50\%.

To achieve file sharing in V1 the Gluster File System \citep{GlusterFS} was used.
The primary and secondary BOINC nodes also had 100GB of EBS attached to each node. 
GlusterFS was configured to replicate every file, so that if the main server failed, the data would not be lost and the system could be restarted as quickly as possible.

\begin{figure}  [hbtp]
\centering
\fbox{\includegraphics [clip, width=0.45\textwidth] {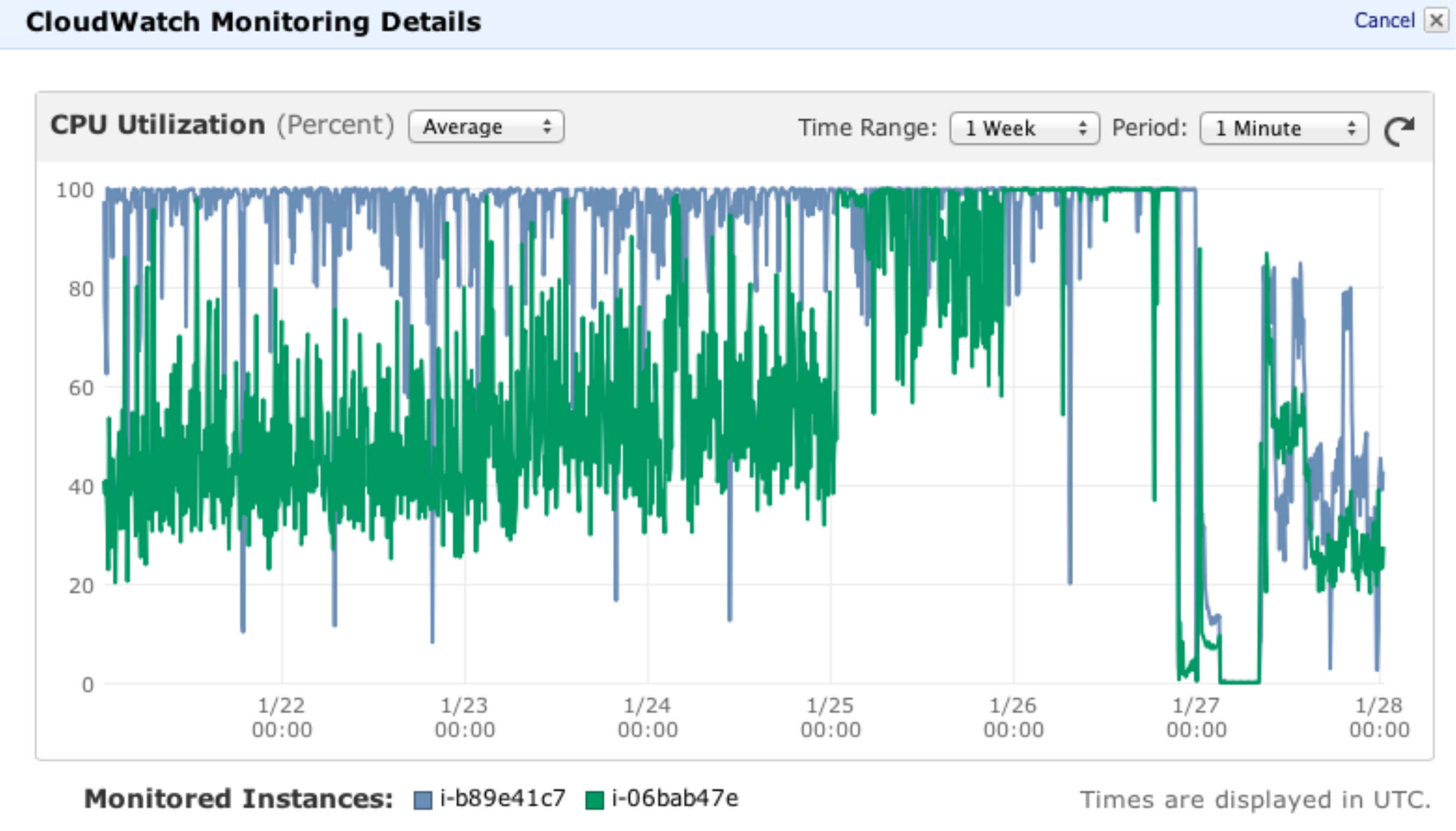}}
\caption{The CPU usage on the AWS servers during the crash in January.}
\label{fig:crash}
\end{figure}

In late January 2013 this was proven in the field, when an EBS failed, due to load, on the primary node during a BOINCstats \citep{boinc-stats2} team challenge.
As can be seen in Figure \ref{fig:crash}, the system was under significant load at the time as the various teams competed to process the most work units.
The node `i-06bab47e' failed around 00:00 UTC on the 27th January, which triggered an alarm.
A new EBS was applied to the server. 
When GlusterFS was restarted the files synchronised from the secondary server and the system was back in operation within 2 hours.
At the same time we moved the EC2 instances from ``small'' to ``medium'' EC2 instances. 

Whilst the GlusterFS was very good it did require quite a lot of maintenance and consumed a significant amount of network bandwidth that we were billed for.
This motivated us to return to the tried and tested NFS server with EBS in a RAID1 configuration.

The AWS model is to charge for data downloaded from the AWS cloud.
In V1 we held the MAGPHYS models on a separate server at The University of Western Australia (UWA). 
The files were marked on the BOINC client as `sticky' and were only downloaded once.
They remained on the client until they were explicitly flushed by the BOINC server.
Amazon have reduced the cost of downloads twice during the course of the project.
A separate model server no longer gives us any benefits so it was removed.

\section{Security}

The BOINC clients allow BOINC software to be run as native code, which could easily compromise the client system. 
The BOINC user community trust the developers not to compromise their machines.
If malicious executables were part of the distribution this would do immense damage to volunteer computing.
As part of the standard BOINC system, all software that is downloaded to the client is signed with a unique certificate generated on the primary BOINC server.
This is to ensure the executables have not been tampered with.

Security on theSkyNet BOINC servers is therefore very important, as they have a shared file system. 
Security patches are applied as soon as they become available.
Access to all the servers is restricted to ssh login via public key authentication. 
Only two people have the privileges to be able to set this up to increase security further.
As shown in Figure \ref{fig:pogs2} many components do not have a public facing IP as they run in a private subnet and therefore cannot be accessed from outside of the VPC.
Thus only a very small number of machines are able to connect to the two public facing theSkyNet servers.

TheSkyNet POGS only makes the BOINC client available from the Berkeley download page \citep{boinc4}.
It is possible for volunteers to build the client on their own machines from the source as BOINC is open source.
If they modify the client to return malicious results, the validator will stop the results continuing as they will not match with other machines.

\section{Web Site}
\label{Web Site}
BOINC provides a PHP based website `out of the box'. 
This includes functionality for running a BOINC project.
Tasks such as:
\begin{enumerate}
\item registering users;
\item forums to discuss things about the project;
\item project statistics;
\item client download;
\item administration tools for the site.
\end{enumerate}

\begin{figure} [hbt]
\centering
\fbox{\includegraphics [clip, width=0.45\textwidth] {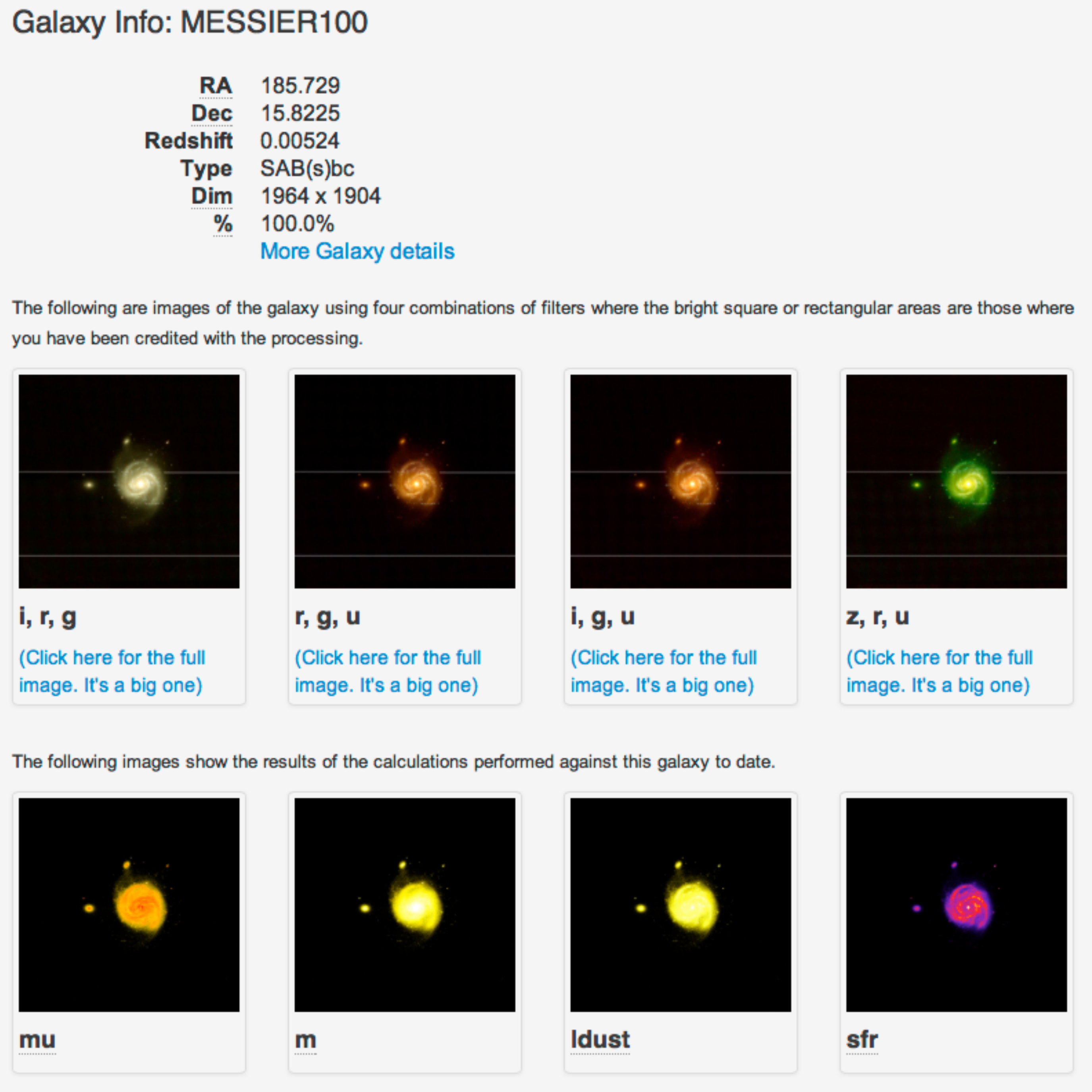}}
\caption{The POGS screen with data for M100.}
\label{fig:pogsImage}
\end{figure}

Under the umbrella of theSkyNet project \citep{theSkyNet}, we have created a separate website.
It allows the users to view the galaxies they have worked on, and to view, in near real time, the state of the SED fitting process any particular galaxy.
The parameters that are displayed have been chosen for their pleasing visual appearance as shown in Figure \ref{fig:pogsImage}. 
The web site also allows the users to win trophies depending on the number of credits earned and see improved statistics about what their computers have been processing.

\begin{figure} [hbt]
\centering
\fbox{\includegraphics [clip, width=0.45\textwidth] {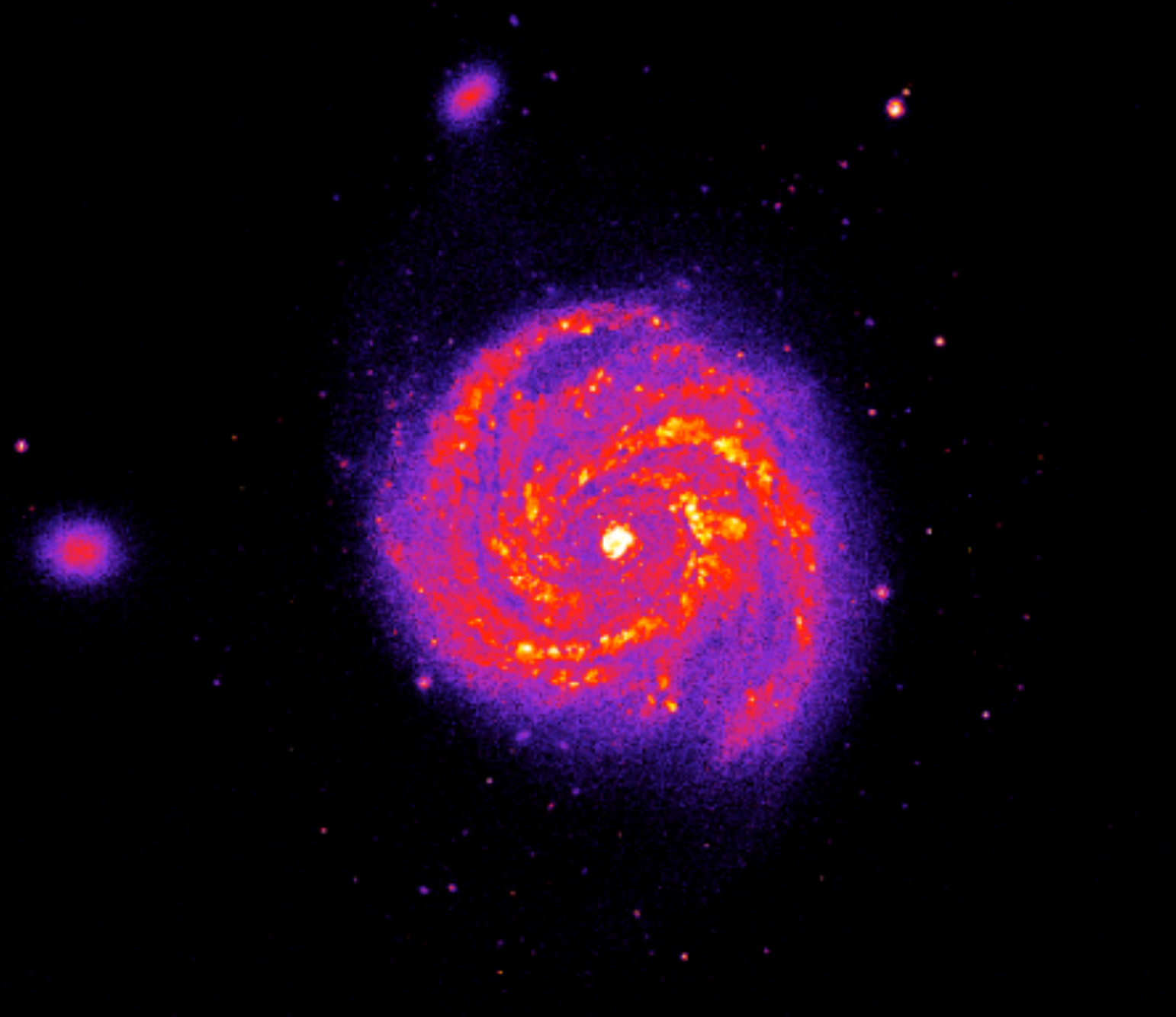}}
\caption{The SFR for M100.}
\label{fig:pogsImage2}
\end{figure}

In Figure \ref{fig:pogsImage}, the top row shows false colour images of the galaxy using a number of different filters.
The bottom row shows false colour images of the current state of various parameters returned by the BOINC clients.
Clicking on any of the images will display a more detailed version of the image as shown in Figure \ref{fig:pogsImage2}. 

\section{Deployment and Testing}
The deployment to AWS is automated as much as possible using: Fabric \citep{Fabric}, Puppet \citep{Puppet} and Boto \citep{Boto}.
The use of AWS, Fabric, Puppet and Boto allow us to create a new software instance within an hour.
The new instance is then automatically saved as an AMI which we can start as and when needed.

This has the significant additional benefit of allowing us to bring up a new test server in a matter of minutes that is an exact replicate of the production system.
A copy of the production database can be created by starting an RDS instance from a snapshot of the production database.
The test server then just requires a few minor configuration changes to point to the new test database. 
Any problematic galaxy can be loaded and tested independently of the main production system.

\section{Performance}
\label{Performance}
\begin{figure}  [hbt]
\centering
\fbox{\includegraphics [clip, width=0.45\textwidth] {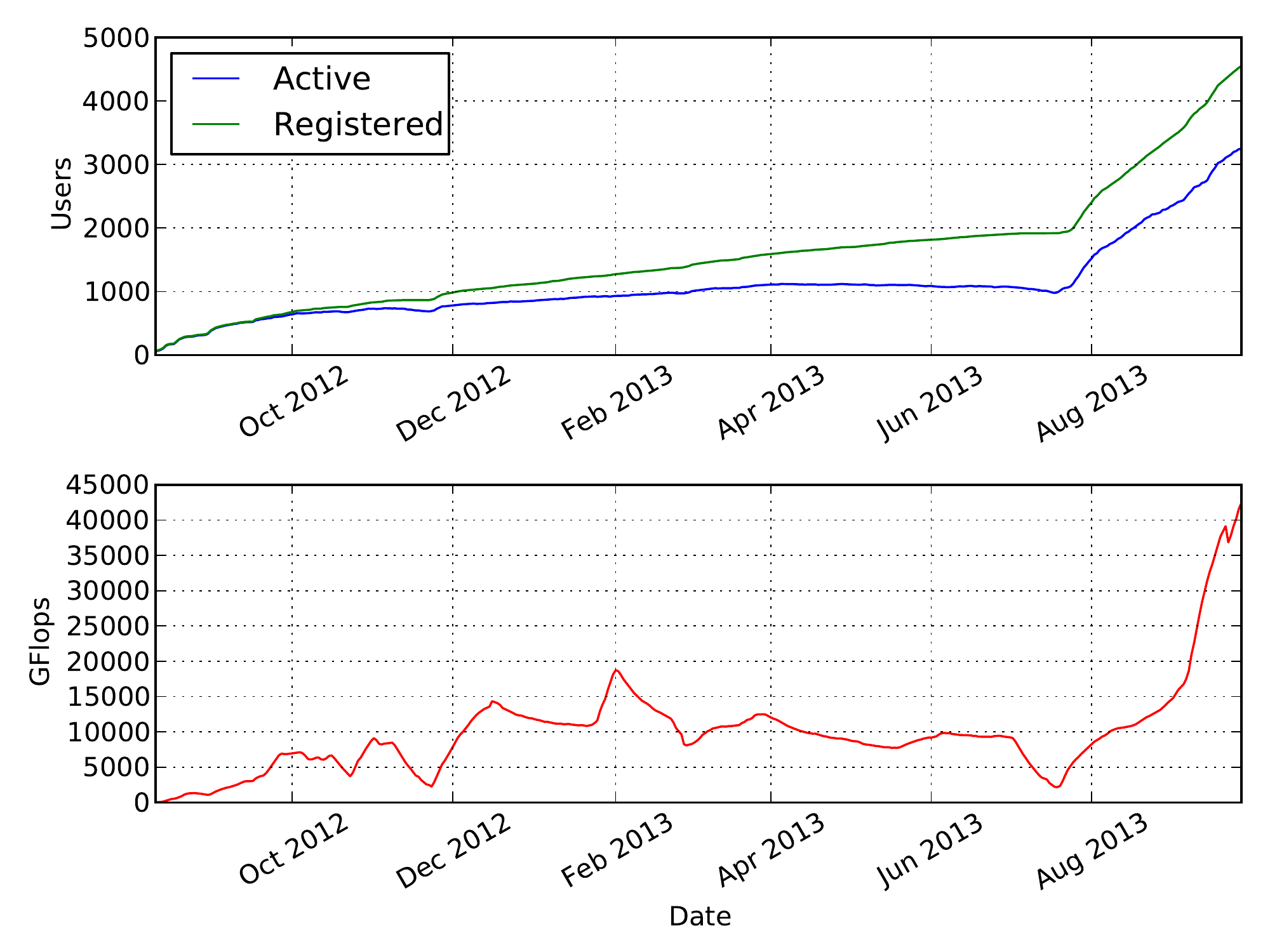}}
\caption{The usage data (as of \asOfDate). 
The upper plot shows the number of active users and registered users.
The lower plot shows an estimate of the GFlops of the clients attached to the POGS system.}
\label{fig:UsageData}
\end{figure}

Figure \ref{fig:UsageData} shows the growth of the POGS project since the BOINC statistics package was activated in August 2012. 
The statistics are reported to sites such as BOINCstats \citep{boinc-stats1} every six hours so the BOINC community can see their credits and their teams credits.
The peak in Feb 2013 was caused by a BOINCstat team challenge~\citep{boinc-stats1}, where 13 teams tried to process as many work units in a two week period as possible.
The peak in Sept 2013 was caused by another BOINCstat team challenge, this time 3 teams tried to process as many work units in a three week period as possible.

The switch from the V1 infrastructure to the V2 happened in mid July 2013 and can be seen by the large dip in performance from the beginning of July.
This was because we did not issue any new work and had to wait for all the issued work units to complete, so we could finish storing the processed galaxies correctly.
The switch to the V2 infrastructure took less than half a day.

After a significant number of records have been inserted and deleted from a MySQL database it is necessary to perform a database optimisation.
Unfortunately this requires a table lock and the database is unusable by the project until it has completed.
The troughs in 2012 and early 2013 are when the database had to be taken offline for maintenance.

The number of registered users are people who have processed work units for the project. 
The number of active users is the number of people who have processed data within the last 3 weeks.
One of the biggest challenges to a BOINC based project is how to retain users.
The volunteers give their CPU cycles freely and need to feel involved in the project.
POGS has offered prizes, which we have mailed to the people who process a particular work unit.
We recently sent prizes to the two people who processed the 2 millionth work unit.

Through theSkyNet project, we plan to undertake a devoted PR campaign to attract users from the amateur astronomical, and the wider BOINC communities.
The ICRAR outreach team have developed strategies to encourage more participation in theSkyNet suite of projects. 
Some of which have been implemented with some good results.

In Figure \ref{fig:UsageData} we can see a sudden influx in users in August and September 2013.
This was due to a number of factors:
\begin{enumerate}
\item in August, an article about the project in The Australian newspaper \citep{Foo:2013}.
\item in August, theSkyNet POGS becoming one of the projects presented when the BOINC client is installed.
\item in September, the second anniversary publicity of theSkyNet.
\item in September, three BOINC challenges running on the project. 
\item in August and September, increased use of social media such as Facebook and Twitter to prompt the project.
\end{enumerate}

\begin{figure}  [hbt]
\centering
\fbox{\includegraphics [clip, width=0.45\textwidth] {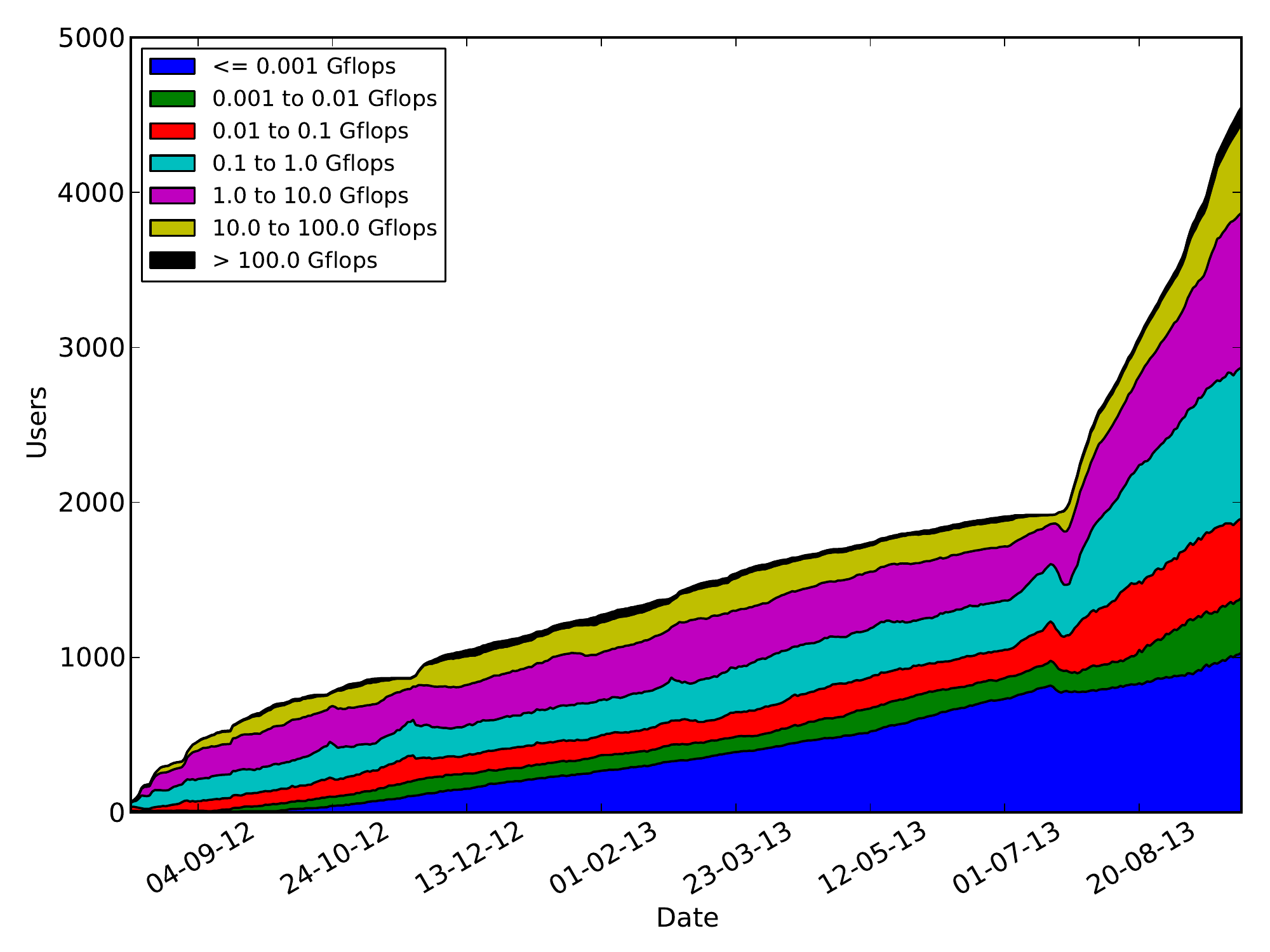}}
\caption{The individual users contributions to the POGS project (as of \asOfDate). 
The x-axis is the date, the y-axis is the number of, and the colour represents the GFlops the band the users are contributing.}
\label{fig:IndividualData}
\end{figure}

Figure \ref{fig:IndividualData} shows the number of individual users by their GFlop contribution, over the course of the project.
As can be seen, many users have assigned slower machines to the project (the large blue area at the bottom), although we are starting to see more mid and high end machines join the project.
We are only now advertising the theSkyNet POGS project and have deliberately let `word of mouth' grow the project for the first year.
This has allowed us to scale the servers gradually to the size we currently need whilst staying within the constraints of the `AWS in Education' grant.
We now know that as the project continues to grow we can easily add more capacity.  

\begin{figure} [hbt]
\centering
\fbox{\includegraphics [clip, width=0.45\textwidth] {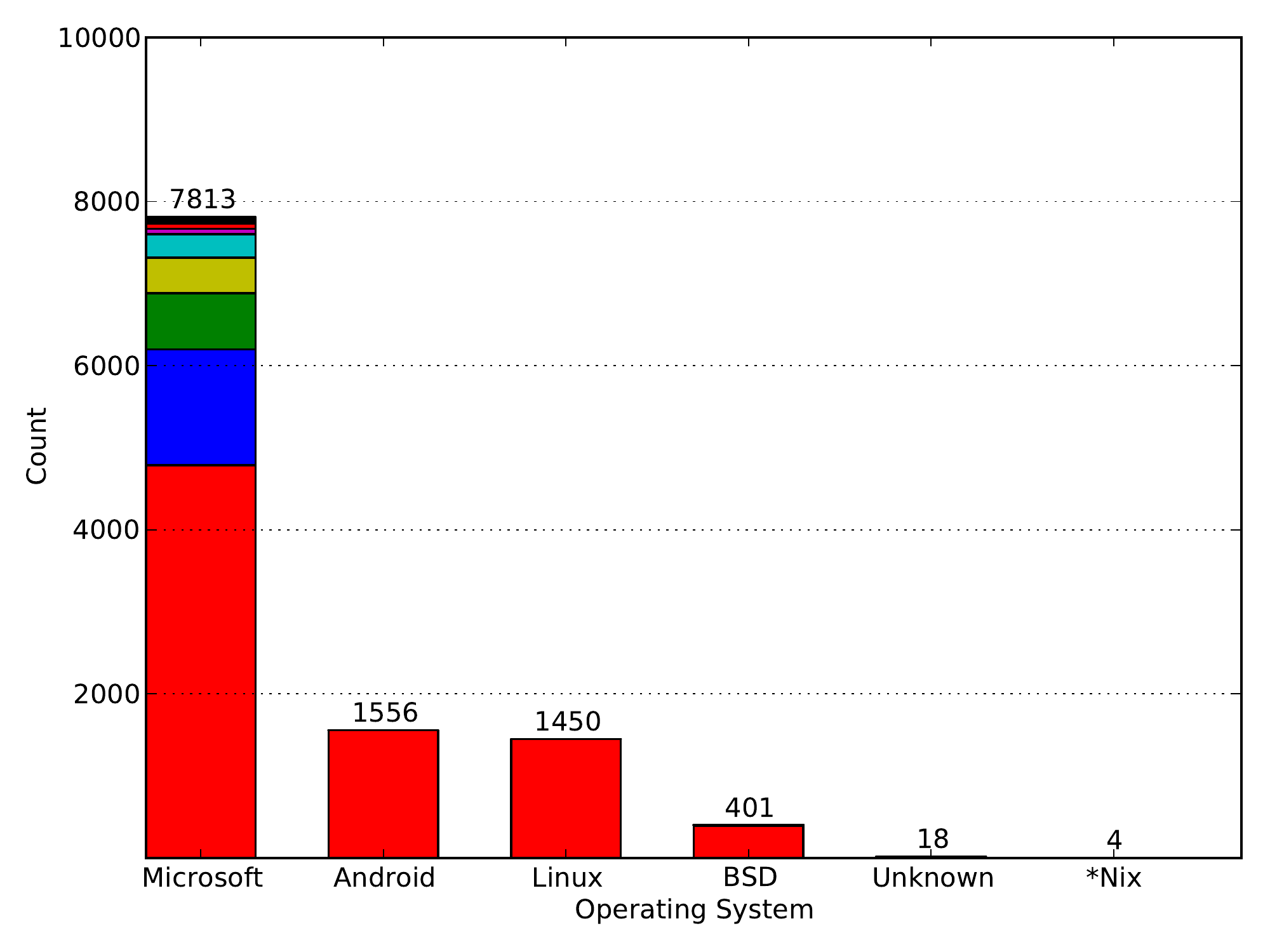}}
\caption{The spread of operating systems (as of \asOfDate). 
Microsoft Windows 7 and XP are the most popular of the 17 flavours of Windows used. OS X dominates the BSD group.}
\label{fig:os_stack}
\end{figure}

\begin{figure} [hbt]
\centering
\fbox{\includegraphics [clip, width=0.45\textwidth] {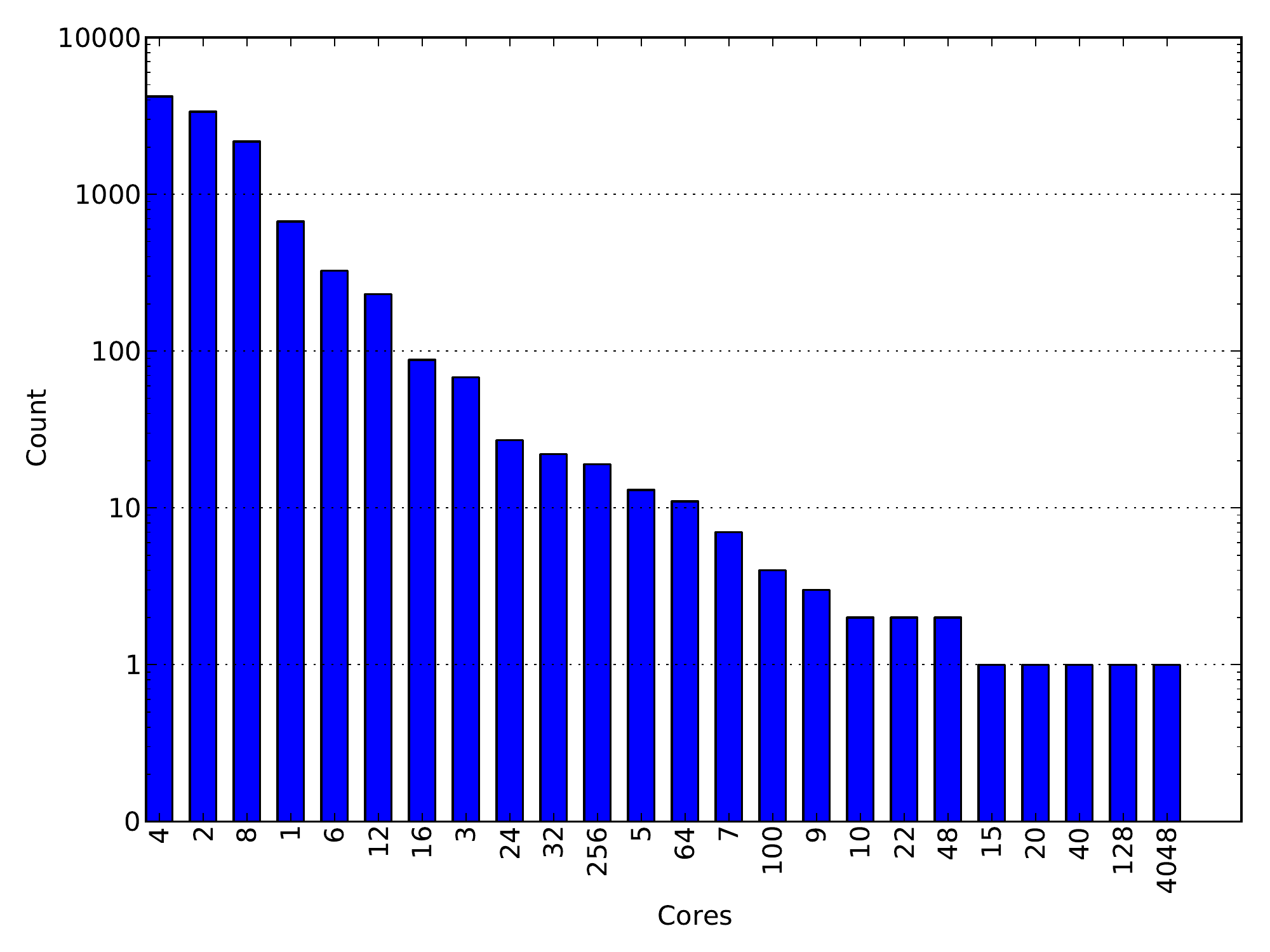}}
\caption{The spread of cores (as of \asOfDate). }
\label{fig:cores}
\end{figure}

The spread of operating system and CPU cores of the various clients are shown in Figure \ref{fig:os_stack} and \ref{fig:cores}. 
Microsoft Windows 7 is the most common operating system.
Interestingly the second and third most common operating system are Android and Linux.
In fourth place is Microsoft Windows XP. 
This is because people have older machines which they dedicate to BOINC processing, whilst keeping their higher performance machines for their personal use.
The project even has 3 Microsoft Windows 98 and 1 Microsoft Millennium machines contributing.

The most popular core types are the quad-core and dual-core processor. 
The quad-core has recently become the most common type.
In the last six months 8-core machines have overtaken single core machines. 
The higher core counts come from Windows server farms.
These appear to the BOINC client as a single node with a large number of cores.

In sections \ref{BOINC distribution & processing} and \ref{Queueing of galaxies} we gave a brief overview of how a BOINC work unit is processed.
Table~\ref{tab:resultsCount} shows that $\sim$90\% of work units match on the first pass with both clients returning a matching result.
The remaining 10\% do not match on the first pass.
This is due to a number of reasons:
\begin{enumerate}
\item issues with the certain Android devices not matching any other machine. We are helping the BOINC community try to debug this issue.
\item the results genuinely did not match.
\item the client did not return the result within the time window.
As most of the BOINC clients work on many projects it is possible for the client to not get round to processing a work unit.
\item the client explicitly aborted the job. 
\item the client had a problem running the software.
\end{enumerate}

\begin{table} [H]
\centering
\begin{tabular}{| l | r | }
\hline
Results & Percentage \\
\hline
\hline
2 & 90.8\% \\
\hline
3 & 8.5\% \\
\hline
4+ & 0.7\% \\
\hline
\end{tabular}
\caption{The number of results per work unit (as of \asOfDate).}
\label{tab:resultsCount}
\end{table}

Table~\ref{tab:resultsOutcome} shows the percentage of the various client connection outcomes. 
As can be seen errors on the client or `no reply' account for most of the errors.
The `Client Error' is primarily people aborting work units so they can be issued to other users.

Table~\ref{tab:resultsValidateState} shows a count of the validation states.
96.2\% of the validations match the first time.

\begin{table} [hbt]
\centering
\begin{tabular}{| l | r | }
\hline
Outcome & Percentage \\
\hline
\hline
Success & 95.6\% \\
\hline
Client error & 3.4\% \\
\hline
No reply & 1.0\% \\
\hline
\end{tabular}
\caption{Outcomes of the various client connections (as of \asOfDate).}
\label{tab:resultsOutcome}
\end{table}

\begin{table} [hbt]
\centering
\begin{tabular}{| l | r | }
\hline
State & Percentage \\
\hline
\hline
Valid    & 96.2\% \\
\hline
Invalid  & 3.7\% \\
\hline
Too late & 0.1\% \\
\hline
\end{tabular}
\caption{The counts of the validate states (as of \asOfDate).}
\label{tab:resultsValidateState}
\end{table}

\section{Conclusions}

TheSkyNet POGS is currently processing approximately 110 galaxies a day.
If we can sustain this the survey would be able to finish the survey in under three years.
The BOINC volunteers appear to be happy to participate in an Astronomy project which puts a significant effort into showing them what their CPU cycles are being used for.

With the permission of \citet{daCunha:2008cy} we have now developed a GPU version of the MAGPHYS code hope to have it deployed for testing in November 2013.
The initial prototype is 5 times faster than the CPU based machine.
The prototype is a straight port of the F77 to OpenCL.
We hope to increase this to 20 times faster than a CPU.

The BOINC middleware has proven to be extremely robust and will be the basis of subsequent theSkyNet projects.
The AWS infrastructure has proven to be trouble free.
As load has increased it has been a relatively simple task to increase the capacity of the servers to handle the load.
Amazon are always reducing the price, which makes it quite an attractive way of trying out systems.
The bill for March 2013 to run the POGS project within AWS was just over AU\$800.
The bill for August 2013 was only just over AU\$450.

Work is currently under way to build some web services at ICRAR to allow people to access the HDF5 data stored in the long term store of iVEC.
The user can get the files as HDF5 or request a subset of data be sent to them in FITS format.

We are starting to work on science outcomes (Thilker et al., in prep.).  
Studies are underway investigating the potential biases that are introduced into results from integrated (total) SED fitting versus the pixel-by-pixel approach.  
With the pixel SED fitting output we also hope to constrain parametric models of galaxy morphology in a more meaningful way than ordinarily achieved. 
In particular, we will fit multi-component (bulge, bar, disk) trial galaxy models directly to the distribution of stellar mass rather than surface brightness in a
single band (which can be hindered by SFH-based variation in the mass-to-light ratio, dust obscuration, and cosmological band-shifting).  
Finally, we are exploring non-parametric measures of galaxy morphology, utilising our physical parameter maps for stellar mass and SFR.  
This effectively adds an orthogonal layer of dimensionality (via SFR) to the widely-used Concentration, Asymmetry, Smoothness, Gini, and M20 system \citep{Conselice:2000jm, Lotz:2004gj}, while also removing single-band biases.

All the source code for this project is available under GNU Lesser General Public License as open source from GitHub \citep{boinc-magphys}.
The main difference between the GPL and the LGPL is that the latter allows the work to be linked with (in the case of a library, `used by') a non-(L)GPLed program, regardless of whether it is free software or proprietary software.
The LGPL allows developers and companies to use and integrate LGPL software into their own (even proprietary) software without being required to release the source code of their own software-parts.


\section{Acknowledgements}

We acknowledge the BOINC Community - who volunteer their time and resources so we can do science.

The Pan-STARRS1 (PS1) Surveys have been made possible through contributions of the Institute for Astronomy, the University of Hawaii, the Pan-STARRS Project Office, the Max-Planck Society and its participating institutes, the Max Planck Institute for Astronomy, Heidelberg and the Max Planck Institute for Extraterrestrial Physics, Garching, The Johns Hopkins University, Durham University, the University of Edinburgh, Queen's University Belfast, the Harvard-Smithsonian Center for Astrophysics, the Las Cumbres Observatory Global Telescope Network Incorporated, the National Central University of Taiwan, the Space Telescope Science Institute, the National Aeronautics and Space Administration under Grant No.\\
NNX08AR22G issued through the Planetary Science Division of the NASA Science Mission Directorate and the University of Maryland.

This publication makes use of data products from the Wide-field Infrared Survey Explorer, which is a joint project of the University of California, Los Angeles, and the Jet Propulsion Laboratory/California Institute of Technology, funded by the National Aeronautics and Space Administration.

This publication makes use of data products based on observations made with the NASA Galaxy Evolution Explorer. 
GALEX is operated for NASA by the California Institute of Technology under NASA contract NAS5-98034.

Funding for the SDSS and SDSS-II has been provided by the Alfred P. Sloan Foundation, the Participating Institutions, the National Science Foundation, the U.S. Department of Energy, the National Aeronautics and Space Administration, the Japanese Monbukagakusho, the Max Planck Society, and the Higher Education Funding Council for England. The SDSS Web Site is http://www.sdss.org/.

The SDSS is managed by the Astrophysical Research Consortium for the Participating Institutions. The Participating Institutions are the American Museum of Natural History, Astrophysical Institute Potsdam, University of Basel, University of Cambridge, Case Western Reserve University, University of Chicago, Drexel University, Fermilab, the Institute for Advanced Study, the Japan Participation Group, Johns Hopkins University, the Joint Institute for Nuclear Astrophysics, the Kavli Institute for Particle Astrophysics and Cosmology, the Korean Scientist Group, the Chinese Academy of Sciences (LAMOST), Los Alamos National Laboratory, the Max-Planck-Institute for Astronomy (MPIA), the Max-Planck-Institute for Astrophysics (MPA), New Mexico State University, Ohio State University, University of Pittsburgh, University of Portsmouth, Princeton University, the United States Naval Observatory, and the University of Washington.

Funding for the NASA-Sloan Atlas has been provided by the NASA Astrophysics Data Analysis Program (08-ADP08-0072) and the NSF (AST-1211644).

This project was initially supported by an Amazon `AWS in Education grant award'.

Finally we'd like to acknowledge Winthrop Professor Gerhardt Meurer and Winthrop Professor Andreas Wicenec for their continued support.

\appendix

\section{Tables}
\begin{table} [H]
\centering
\begin{tabular}{ | l | L{0.35\textwidth} | }
\hline
Parameter & Description of MAGPHYS physical parameter \\
\hline
\hline
$f_\mu^{SFH}$ & the fraction of total infrared luminosity contributed by dust in the ambient ISM from the stellar population spectrum model \\
\hline
$f_\mu^{IR}$ & the fraction of total infrared luminosity contributed by dust in the ambient ISM from the infrared spectrum model \\
\hline
$\mu$ & the fraction of the total effective V-band absorption optical depth of the dust contributed by the ambient ISM \\
\hline
$\hat{\tau}_V$ & the total effective V-band absorption optical depth of the dust seen by young stars inside birth clouds \\
\hline
$\psi_S$ & the specific star formation rate \\
\hline
$M_*$ & the stellar mass \\
\hline
$L_d^{tot}$ & the total luminosity absorbed and reradiated by dust \\
\hline
$T_C^{ISM}$ & the characteristic temperature of cold dust grains which can only reside in the ambient interstellar medium \\
\hline
$T_W^{BC}$ & the characteristic temperature of warm dust grains residing in stellar birth clouds \\
\hline
$\xi_C^{tot}$ & the total contribution of cold dust grains to the total infrared luminosity \\
\hline
$\xi_{PAH}^{tot}$ & the total contribution of Polycyclic Aromatic Hydrocarbons (PAH) to the total infrared luminosity \\
\hline
$\xi_{MIR}^{tot}$ & the total contribution of mid-infrared continuum emissions to the total infrared luminosity \\
\hline
$\xi_W^{tot}$ & the total contribution of warm dust grains to the total infrared luminosity \\
\hline
$\mu\hat{\tau}_V$ & the effective V-band absorption optical depth of the dust in the ambient ISM \\
\hline
$M_d$ & the total dust mass \\
\hline
$\psi(t)$ & the star formation rate \\
\hline
\end{tabular}
\caption{The parameters calculated by MAGPHYS and stored in the database.}
\label{tab:magphys}
\end{table}

\begin{table} [H]
\centering
\begin{tabular}{ | l | p{0.35\textwidth} | }
\hline
Number & Description of stored value \\
\hline
\hline
1 & Best Fit \\
\hline
2 & Median - 50th percentile of the parameter distribution \\
\hline
3 & Highest Probability Bin \\
\hline
4 & 2.5th percentile of the parameter distribution \\
\hline
5 & 16th percentile of the parameter distribution \\
\hline
6 & 84th percentile of the parameter distribution \\
\hline
7 & 97.5th percentile of the parameter distribution \\
\hline
\end{tabular}
\caption{The values calculated for each of the MAGPHYS parameters.}
\label{tab:magphys-layers}
\end{table}


\section{References}

\bibliographystyle{model2-names}
\bibliography{astronomy-and-computing}







\end{document}